\documentclass[12pt,english,floatfix,nofootinbib,superscriptaddress,aps,prd,preprint]{revtex4}
\usepackage[utf8]{inputenc}
\usepackage{float}
\usepackage{array}
\usepackage{bbold}
\usepackage{lipsum}
\usepackage{dsfont}
\usepackage{graphicx}
\usepackage{amsmath,amsthm,amsfonts,amssymb}
\usepackage{graphicx}
\usepackage[english]{babel} 
\usepackage{color}
\usepackage{tensor}
\usepackage{esint}
\usepackage[dvips]{epsfig}
\usepackage[dvips]{graphicx}
\usepackage{float}
\usepackage{units}
\usepackage{textcomp}
\usepackage{mathrsfs}
\usepackage{amsmath}
\usepackage[makeroom]{cancel}
\usepackage{amssymb}
\usepackage{amsbsy}
\usepackage{amsfonts}
\usepackage{amssymb,mathrsfs,xcolor}
\usepackage{esint}
\usepackage{braket}
\usepackage{array}
\usepackage{graphicx}

\usepackage{wasysym}
\usepackage{multirow}
\usepackage{wrapfig}
\usepackage{subfig}

\usepackage{stmaryrd}
\usepackage{upgreek}

\makeatletter

\makeatletter\usepackage{babel}

\usepackage{hyperref}
\hypersetup{
    colorlinks,
    citecolor=blue,
    filecolor=green,
    linkcolor=purple,
    urlcolor=red,
}

\usepackage{slashed}

\newcommand{\ie}{\begin{equation}}
\newcommand{\fe}{\end{equation}}
\newcommand{\se}{\begin{eqnarray}}
\newcommand{\ff}{\end{eqnarray}}

\begin{document}

\title{Gravitational traces of bumblebee gravity in metric--affine formalism}

\author{A. A. Ara\'{u}jo Filho}
\email{dilto@fisica.ufc.br}

\affiliation{Departamento de Física Teórica and IFIC, Centro Mixto Universidad de Valencia--CSIC. Universidad
de Valencia, Burjassot--46100, Valencia, Spain}

\affiliation{Departamento de Física, Universidade Federal da Paraíba, Caixa Postal 5008, 58051-970, João Pessoa, Paraíba,  Brazil.}

\author{H. Hassanabadi}
\email{hha1349@gmail.com}
\affiliation{Department of Physics, University of Hradec Kr$\acute{a}$lov$\acute{e}$, Rokitansk$\acute{e}$ho 62, 500 03 Hradec Kr$\acute{a}$lov$\acute{e}$, Czechia.}
\affiliation{Physics Department, Shahrood University of Technology, Shahrood, Iran}

\author{N. Heidari}
\email{heidari.n@gmail.com}

\affiliation{Physics Department, Shahrood University of Technology, Shahrood, Iran}

\author{J. K\u{r}\'\i\u{z}}
\email{jan.kriz@uhk.cz}

\affiliation{Department of Physics, University of Hradec Kr$\acute{a}$lov$\acute{e}$, Rokitansk$\acute{e}$ho 62, 500 03 Hradec Kr$\acute{a}$lov$\acute{e}$, Czechia.}

\author{S. Zare}
\email{soroushzrg@gmail.com} 

\affiliation{Department of Physics, University of Hradec Kr$\acute{a}$lov$\acute{e}$, Rokitansk$\acute{e}$ho 62, 500 03 Hradec Kr$\acute{a}$lov$\acute{e}$, Czechia.}


\begin{abstract}

This work explores various manifestations of bumblebee gravity within the metric--affine formalism. We investigate the impact of the Lorentz violation parameter, denoted as $X$, on the modification of the \textit{Hawking} temperature. Our calculations reveal that as $X$ increases, the values of the \textit{Hawking} temperature attenuate. To examine the behavior of massless scalar perturbations, specifically the \textit{quasinormal} modes, we employ the WKB method. The transmission and reflection coefficients are determined through our calculations. The outcomes indicate that a stronger Lorentz--violating parameter results in slower damping oscillations of gravitational waves. To comprehend the influence of the \textit{quasinormal} spectrum on time--dependent scattering phenomena, we present a detailed analysis of scalar perturbations in the time--domain solution. Additionally, we conduct an investigation on shadows, revealing that larger values of $X$ correspond to larger shadow radii. Furthermore, we constrain the magnitude of the shadow radii using the EHT horizon--scale image of $Sgr A^*$. Finally, we calculate both the time delay and the deflection angle.

\end{abstract}

\maketitle


\bigskip

\newpage
\section{Introduction}

The problem of consistently implementing Lorentz symmetry breaking (LSB) within the gravitational framework is fundamentally different from constructing Lorentz--breaking extensions for non--gravitational field theories. While flat spacetimes can accommodate Lorentz--breaking additive terms like Carroll--Field--Jackiw \cite{1}, aether time \cite{2}, and others (see, for example, \cite{3}), these constructions cannot be readily applied to curved spacetimes.

In flat spacetime, constant tensors can be well--defined, allowing for simple conditions such as $\partial_{\mu}k_{\nu}=0$. However, in curved spacetimes, these conditions cannot be introduced in an analogous manner. The non--covariant condition $\partial_{\mu}k_{\nu}=0$ is not suitable, and its covariant extension $\nabla_{\mu}k_{\nu}=0$ imposes severe restrictions on spacetime geometries (known as the no--go constraints \cite{4}). Therefore, the most natural way to incorporate local Lorentz violation into gravitational theories is through the mechanism of spontaneous symmetry breaking. In this case, Lorentz/CPT violating coefficients (operators) arise as vacuum expectation values (VEV) of dynamical tensor fields, which are driven by nontrivial potentials.

The generic effective field framework that describes all possible coefficients for Lorentz/CPT violation is the well--known Standard Model Extension (SME) \cite{5}. The gravitational sector of SME is defined in a Riemann--Cartan manifold, where the torsion is treated as a dynamic geometric quantity alongside the metric. However, despite the non--Riemannian background, most studies in bumblebee gravity, for example, have been conducted within the metric approach of gravity, where the metric is the only dynamical geometric field. These investigations focus on obtaining exact solutions for different models that incorporate LSB in curved spacetimes, including bumblebee gravity \cite{6,7,8,9,10,11,12,13}, parity--violating models \cite{14,15,16,17,18}, and Chern--Simons modified gravity \cite{19,20,21}.

While the majority of works in the literature on modified theories of gravity employ the metric approach, it is interesting to explore more general geometric frameworks. Motivations for considering theories of gravity in Riemann--Cartan background include, allowing for the possibility of having gravitational topological terms \cite{22}. Another intriguing non--Riemannian geometry that has garnered attention is Finsler geometry \cite{23}, which has been linked to LSB in recent years through various studies \cite{24,25,26,27,28}.

However, the most compelling generalization of the metric approach is the metric--affine formalism, where the metric and connection are treated as independent dynamic geometric quantities. Despite this formalism significance, LSB remains relatively unexplored in the literature. Recent works involving bumblebee gravity have started to address this gap within the metric--affine approach \cite{31,32,33,34,nascimento2022vacuum}.

Understanding gravitational waves and their characteristics is essential for studying various physical processes, including cosmological events in the early universe and astrophysical phenomena like the evolution of stellar oscillations \cite{unno1979nonradial, kjeldsen1994amplitudes,dziembowski1992effects} and binary systems \cite{pretorius2005evolution,hurley2002evolution,yakut2005evolution,heuvel2011compact}. Gravitational waves exhibit a range of intensities and characteristic modes, and their spectral properties depend on the phenomena that generate them \cite{riles2017recent}. Of particular interest is the emission of gravitational waves from black holes. When a black hole forms through the gravitational collapse of matter, it enters a perturbed state and emits radiation that includes a bundle of characteristic frequencies unrelated to the collapse process \cite{konoplya2011quasinormal}. These perturbations with distinct frequencies are referred to as \textit{quasinormal} modes \cite{heidari2023gravitational,kokkotas1999quasi}.

The investigation of quasinormal modes of black holes has extensively utilized the weak field approximation in the literature, both in general relativity (GR) \cite{rincon2020greybody,santos2016quasinormal,oliveira2019quasinormal,berti2009quasinormal,horowitz2000quasinormal,nollert1999quasinormal,ferrari1984new,kokkotas1999quasi,london2014modeling,maggiore2008physical,flachi2013quasinormal,ovgun2018quasinormal,blazquez2018scalar,roy2020revisiting,konoplya2011quasinormal} and in alternative gravity theories such as Ricci--based theories \cite{kim2018quasi,lee2020quasi,jawad2020quasinormal}, Lorentz violation \cite{maluf2013matter,maluf2014einstein}, and related fields \cite{JCAP1,JCAP2,JCAP3,jcap4,jcap5}.

Significant progress has been made in the development of gravitational wave detectors, enabling the detection of gravitational waves emitted from various physical phenomena \cite{abbott2016ligo, abbott2017gravitational, abbott2017gw170817, abbott2017multi}. Ground--based interferometers like VIRGO, LIGO, TAMA--300, and EO--600 have played crucial roles in these detections \cite{fafone2015advanced,abramovici1992ligo,coccia1995gravitational,luck1997geo600}. Over the years, these detectors have improved their accuracy, approaching the desired sensitivity range \cite{evans2014gravitational}. These detections have provided valuable insights into the composition of astrophysical objects, including boson stars and neutron stars.

The detection of gravitational waves has a significant connection to black hole physics. The emitted gravitational radiation from a perturbed black hole carries a unique signature that allows for direct observations of their existence \cite{thorne2000probing}. Early studies on black hole perturbations were conducted by Regge and Wheeler, who examined the stability of Schwarzschild black holes \cite{regge1957stability}, followed by Zerilli's work on perturbations \cite{zerilli1970effective,zerilli1974perturbation}.

The study of gravitational solutions involving scalar fields has received significant attention in recent years due to their notable and peculiar characteristics. These intriguing aspects have led to the emergence of various astrophysical applications. Notably, black holes (BHs) with nontrivial scalar fields challenge the well--known ``no--hair theorem" \cite{herdeiro2015asymptotically}. Additionally, the existence of long--lived scalar field patterns \cite{ayon2016analytic}, the formation of boson stars \cite{colpi1986boson,palenzuela2017gravitational,cunha2017lensing}, and the exploration of exotic astrophysical scenarios in the form of gravastars \cite{visser2004stable,pani2009gravitational,chirenti2016did} have all stemmed from this line of research. Moreover, by considering Klein--Gordon scalar fields on curved backgrounds, a wide range of phenomena can be derived, including black hole bombs \cite{cardoso2004black,sanchis2016explosion,hod2016charged} and superradiance \cite{brito2015black}.

In this study, we thoroughly investigate bumblebee gravity within the metric--affine formalism, examining its diverse manifestations and shedding light on its intriguing properties. Our analysis encompasses the modification of the \textit{Hawking} temperature, the behavior of \textit{quasinormal} modes, the influence on time--dependent scattering phenomena, the analysis of shadows, the exploration of time delay. We reveal the correlation between bumblebee gravity and thermodynamic properties, with the \textit{Hawking} temperature attenuating as the Lorentz violation parameter increases. Furthermore, our investigation highlights the distinct influence of bumblebee field on gravitational wave damping, with slower oscillations observed under stronger Lorentz violation. The analysis of shadows provides valuable insights, demonstrating larger shadow radii associated with higher values of the parameter. Additionally, we provide the analysis of the time delay in this scenario.


\section{The general setup and Hawking temperature}
\label{general}

In this section, we begin by introducing a metric--affine generalization of the gravitational sector found in the SME \cite{5}. Like the metric formulation, the action of this sector can be expressed in the following manner
\begin{eqnarray}
    \nonumber\mathcal{S} &=& \frac{1}{2 \kappa^{2}} \int \mathrm{d}^{4}x \sqrt{-g} \left\{ (1-u)R(\Gamma) + s^{\mu\nu}R_{\mu\nu}(\Gamma) + t^{\mu\nu\alpha\beta}R_{\mu\nu\alpha\beta}(\Gamma)    \right\}+\mathcal{S}_{mat}(g_{\mu\nu},\psi)+\\
    &+& \mathcal{S}_{coe}(g_{\mu\nu},u,s^{\mu\nu},t^{\mu\nu\alpha\beta}).
    \label{1}
\end{eqnarray}
Here, we have $g_{\mu\nu}$ as the covariant metric tensor, $g^{\mu\nu}$ as the contravariant metric tensor, and the determinant of the metric is given by $g=\text{det}(g_{\mu\nu})$.
In this formulation, we express the action as follows, incorporating various geometric quantities. The Ricci scalar $R(\Gamma)\equiv g^{\mu\nu}R_{\mu\nu}(\Gamma)$, Ricci tensor $R_{\mu\nu}(\Gamma)$, and Riemann tensor $R^{\mu}_{\,\,\,\nu\alpha\beta}(\Gamma)$ play key roles. Additionally, $\mathcal{S}_{mat}$ represents the action that accounts for the contributions of matter sources, which are assumed to be coupled solely to the metric\footnote{It is worth noting that fermions naturally couple to the connection. However, for convenience, we omit consideration of spinors in this discussion and focus solely on bosonic matter sources that are minimally coupled to the metric.}. As previously mentioned, we adopt the metric--affine  formalism, wherein the metric and connection are regarded as independent dynamical entities {\it a priori}.

Moreover, the coefficients (fields) $u=u(x)$, $s^{\mu\nu}=s^{\mu\nu}(x)$, and $t^{\mu\nu\alpha\beta}=t^{\mu\nu\alpha\beta}(x)$ play a crucial role in explicitly breaking local Lorentz symmetry, as extensively discussed in Ref. \cite{5}. It is important to note that the background field $s^{\mu\nu}$ possesses the same symmetries as the Ricci tensor. However, for the purposes of this study, we assume it to be a symmetric second--rank tensor, denoted as $s^{\mu\nu}=s^{(\mu\nu)}$. Therefore, it couples exclusively to the symmetric component of the Ricci tensor. On the other hand, $t^{\mu\nu\alpha\beta}$ shares the symmetries of the Riemann tensor. Lastly, the final term in Eq. (\ref{1}), denoted as $\mathcal{S}_{coe}$, encompasses the dynamical contributions stemming from the Lorentz--violating coefficients.

In this study, our primary focus lies on investigating the nontrivial consequences of Lorentz symmetry breaking, which involve both the Ricci tensor and scalar. Consequently, we impose constraints such that the coefficients $s^{\mu\nu}$ and $u$ take nonzero values, while setting $t^{\mu\nu\alpha\beta} = 0$. This choice is motivated by the fact that the connection equation cannot be straightforwardly solved through a simple metric redefinition when a nontrivial parameter $t^{\mu\nu\alpha\beta}$ is present. This issue is commonly referred to as the ``t--puzzle" \cite{35,maluf2019antisymmetric}. Thus, the action of interest can be expressed as follows:
\begin{equation}
\mathcal{S} = \frac{1}{2 \kappa^{2}} \int \mathrm{d}^{4}x \sqrt{-g} \left\{ (1-u)R(\Gamma) + s^{\mu\nu}R_{\mu\nu}(\Gamma)    \right\}+\mathcal{S}_{mat}+\mathcal{S}_{coe}.
\label{S2}
\end{equation}
It is important to emphasize that the aforementioned action exhibits invariance under projective transformations of the connection
\begin{equation}
\Gamma^{\mu}_{\nu\alpha}\longrightarrow \Gamma^{\mu}_{\nu\alpha}+\delta^{\mu}_{\alpha}A_{\nu}.
\label{Proj}
\end{equation}
Under the this transformation described by Eq. (\ref{Proj}), where $A_{\alpha}$ represents an arbitrary vector, it can be readily verified that the Riemann tensor undergoes the following change:
\begin{equation}
    R^{\mu}_{\,\,\,\nu\alpha\beta}\longrightarrow R^{\mu}_{\,\,\,\nu\alpha\beta}-2\delta^{\mu}_{\nu}\partial_{[\alpha}A_{\beta]}.
\end{equation}
Consequently, it follows that the symmetric part of the Ricci tensor remains invariant under the transformation described by Eq. (\ref{Proj}), along with the entire action (\ref{S2}). The model presented by the action (\ref{S2}) belongs to a broader class of gravitational theories known as Ricci--based theories \cite{36,37}. It has been demonstrated that within this class of models, the property of projective invariance prevents the appearance of ghost--like propagating degrees of freedom in gravity.

The static and spherically symmetric solution in metric--affine traceless bumblebee model is given by \cite{nascimento2022vacuum}
\begin{equation}
    \mathrm{d}s^2_{(g)}=-\frac{\left(1-\frac{2M}{r}\right)\mathrm{d}t^2}{\sqrt{\left(1+\frac{3X}{4}\right)\left(1-\frac{X}{4}\right)}}+\frac{\mathrm{d}r^2}{\left(1-\frac{2M}{r}\right)}\sqrt{\frac{\left(1+\frac{3X}{4}\right)}{\left(1-\frac{X}{4}\right)^3}}+r^{2}\left(\mathrm{d}\theta^2 +\sin^{2}{\theta}\mathrm{d}\phi^2\right),
    \label{metric3}
\end{equation}
where we have used the shorthand notation: $X=\tilde{\xi} b^2$, $b_{\mu}$ is the vacuum expectation value of the bumblebee field $B_{\mu}$, i.e., $<B_{\mu}>=b_{\mu}$, and $\tilde{\xi}$ is a dimensionless parameter. Note that the line element in Eq. (\ref{metric3}) describes a LSB modified Schwarzschild metric.

The historical foundations of the thermodynamic interpretation of gravity trace back to the pioneering research of the mid--1970s, notably by Bekenstein \cite{bekenstein1973black, bekenstein1974generalized, bekenstein2020black} and Hawking \cite{bardeen1973four} on the thermodynamics of black holes. 
In 1995, Jacobson made a significant advancement in our understanding by demonstrating that the Einstein field equations, the fundamental mathematical framework characterizing relativistic gravitation, can be derived from the amalgamation of overarching thermodynamic principles and the foundational concept of the equivalence principle \cite{jacobson1995thermodynamics}. This groundbreaking revelation marked a pivotal milestone in the synthesis of gravitational theory and thermodynamics. Subsequent to this critical research, Padmanabhan further explored the link between gravity and the concept of entropy \cite{padmanabhan2010thermodynamical}.

\begin{figure}
    \centering
    \includegraphics[scale=0.5]{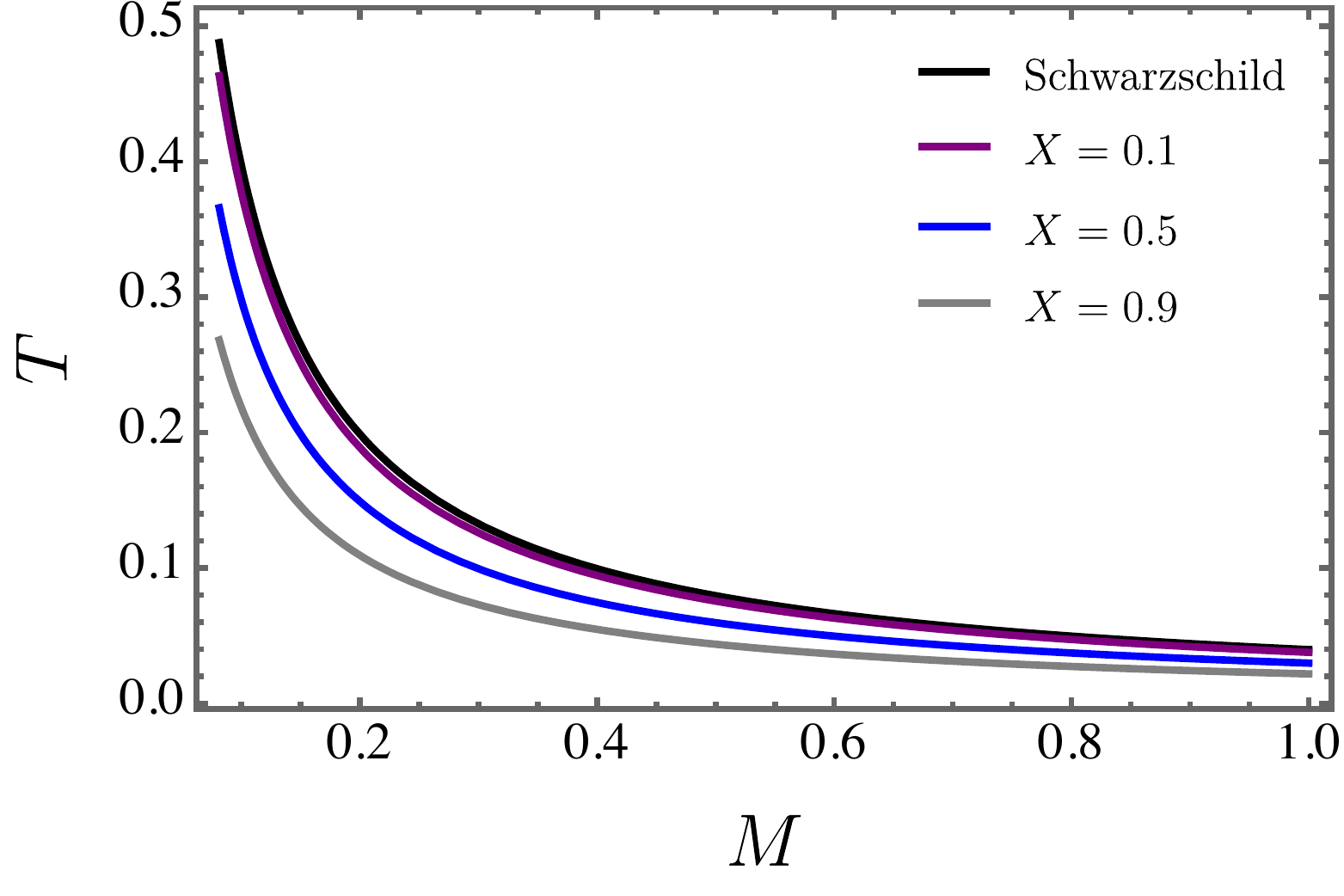}
    \caption{The modifications of Hawking temperature caused by different values of $X$.}
    \label{modifiedtemperature}
\end{figure}

In the context of the thermodynamic properties, the unique geometric quantity that will be affected is the \textit{Hawking} temperature
\begin{equation}\label{R0}
\ {T } = \frac{1}{{4\pi \sqrt {{g_{00}}{g_{11}}} }}{\left. {\frac{{\mathrm{d}{g_{00}}}}{{\mathrm{d}r}}} \right|_{r = {r_{s}}}} \approx \frac{1}{8 \pi  M}-\frac{X}{16 (\pi  M)},
\end{equation}
where $r_{s}$ is the Schwarzschild radius. This thermal quantity, as shown in Fig. \ref{modifiedtemperature}, illustrates the relationship between Lorentz--violating parameters and the respective temperatures. It is evident that as $X$ increases, the corresponding \textit{Hawking} temperatures become increasingly attenuated. It is important to mention that the thermodynamic aspects were calculated for a variety contexts, including cosmological scenarios \cite{araujo2021bouncing,araujo2022thermal,araujo2023thermodynamics,aa2,aa4,aa6,aa7,aa10,aa13,aa14,aa15,filho2022thermodynamics} and more \cite{aa11,aa12,aa1,qq}.


\section{The quasinormal modes}

During the ringdown phase, an intriguing phenomenon known as \textit{quasinormal} modes becomes evident. These modes exhibit characteristic oscillation patterns that are independent of the initial perturbations and instead reflect the intrinsic properties of the system. This remarkable behavior arises from the fact that the \textit{quasinormal} modes correspond to the free oscillations of spacetime itself, unaffected by the specific initial conditions.

In contrast to the \textit{normal} modes, which are associated with closed systems, the \textit{quasinormal} modes are linked to open systems. Consequently, these modes gradually lose energy by radiating gravitational waves. Mathematically, the \textit{quasinormal} modes can be described as the poles of the complex Green function.

To determine the frequencies of the \textit{quasinormal} modes, we need to find solutions to the wave equation for a system described by a background metric $g_{\mu\nu}$. However, analytical solutions for such modes are generally unattainable.

Numerous techniques have been proposed in scholarly works to derive solutions for the \textit{quasinormal} modes. Among these methods, the WKB (Wentzel--Kramers--Brillouin) approach stands out as one of the most renowned. Its inception can be traced back to Will and Iyer \cite{iyer1987black,iyer1987black1}, and subsequent advancements up to the sixth order were made by Konoplya \cite{konoplya2003quasinormal}. For our calculations, we specifically focus on investigating perturbations utilizing the scalar field. Consequently, we formulate the Klein--Gordon equation in the context of a curved spacetime
\ie
\frac{1}{\sqrt{-g}}\partial_{\mu}(g^{\mu\nu}\sqrt{-g}\partial_{\nu}\Phi) = 0.\label{KL}
\fe

While the exploration of \textit{backreaction} effects is intriguing in this particular scenario, this manuscript focuses on other aspects and does not delve into this feature. Instead, our attention is directed towards studying the scalar field as a minor perturbation. Furthermore, given the presence of spherical symmetry, we can exploit the opportunity to decompose the scalar field in a specific manner, as outlined below:
\ie
\Phi(t,r,\theta,\varphi) = \sum^{\infty}_{l=0}\sum^{l}_{m=-l}r^{-1}\Psi_{lm}(t,r)Y_{lm}(\theta,\varphi),\label{decomposition}
\fe
where the spherical harmonics are represented by $Y_{lm}(\theta,\varphi)$. By substituting the scalar field decomposition, as given in Eq. (\ref{decomposition}), into Eq. (\ref{KL}), we arrive at a Schrödinger--like equation. This transformed equation exhibits wave--like characteristics, making it suitable for our analytical investigation
\ie
-\frac{\partial^{2} \Psi}{\partial t^{2}}+\frac{\partial^{2} \Psi}{\partial r^{*2}} + V_{eff}(r^{*})\Psi = 0.\label{schordingereq}
\fe
The potential $V_{eff}$ is commonly referred to as the \textit{Regge--Wheeler} potential, also known as the effective potential. It contains valuable information regarding the geometry of the black hole. Additionally, we introduce the tortoise coordinate $r^{*}$ (running $r^{*}\rightarrow \pm \infty$ all over the spacetime) as $\mathrm{d} r^{*} = \sqrt{-g_{11}/g_{00}}\,\mathrm{d}r$, which reads
\ie
\begin{split}
r^{*}= \sqrt{\frac{\left( 1+\frac{3X}{4} \right)}{\left( 1-\frac{X}{4} \right)}} \left(r + 2M \ln |r-2M| \right).
\end{split}
\fe
After performing several algebraic manipulations, we can express the effective potential explicitly as follows:
\ie
\begin{split}
V_{eff}(r) = \left( 1 - \frac{2M}{r} \right) \left[ \frac{\left(  1-\frac{X}{4} \right)}{\left(1+\frac{3X}{4} \right)} \frac{2M}{r^{3}}  + \frac{l(l+1)}{r^{2} \sqrt{\left(1+\frac{3X}{4} \right)  \left(  1-\frac{X}{4} \right)      }}\right]
\end{split}.
\fe

\begin{figure}
    \centering
    \includegraphics[scale=0.365]{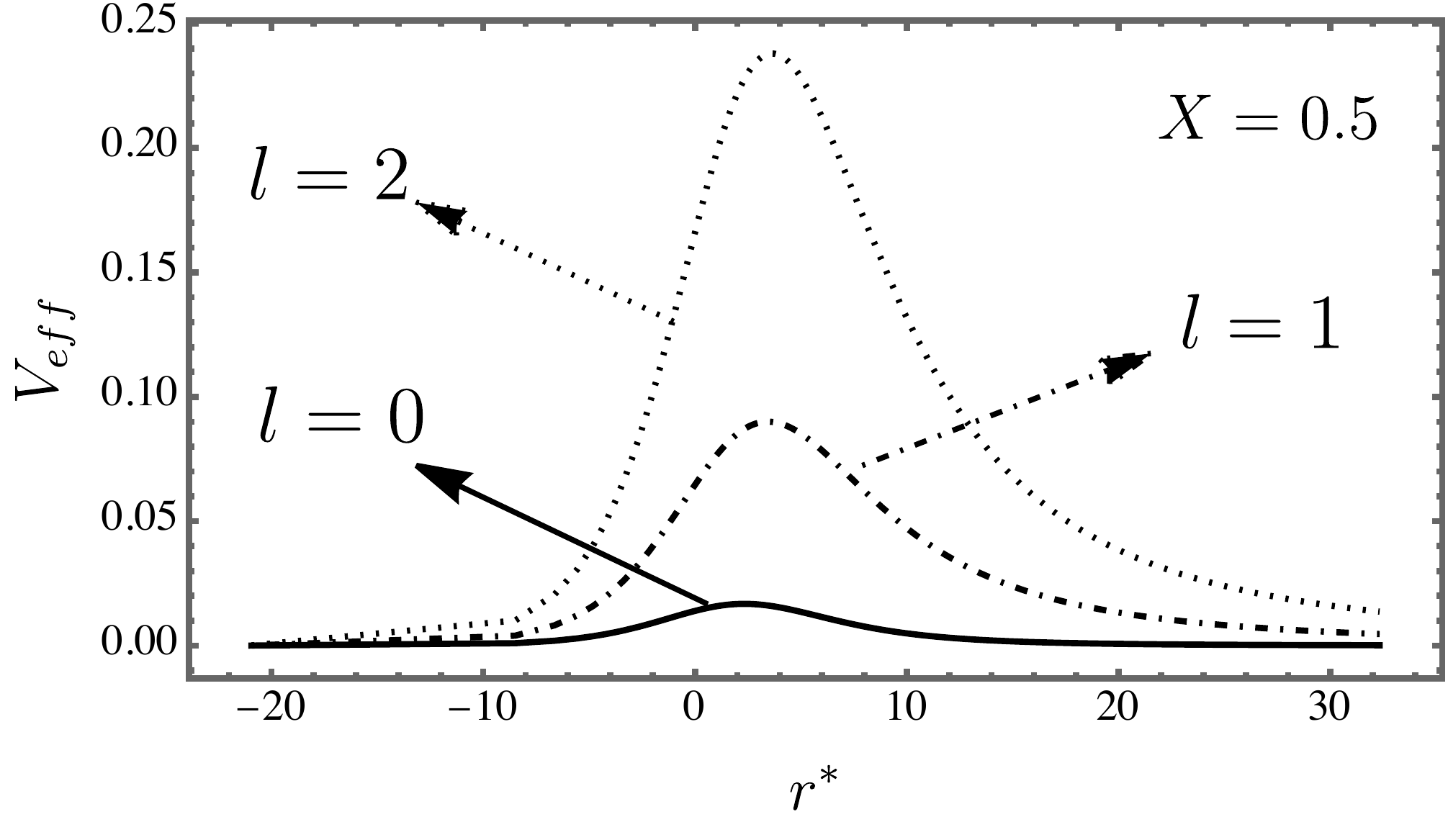}
    \includegraphics[scale=0.4]{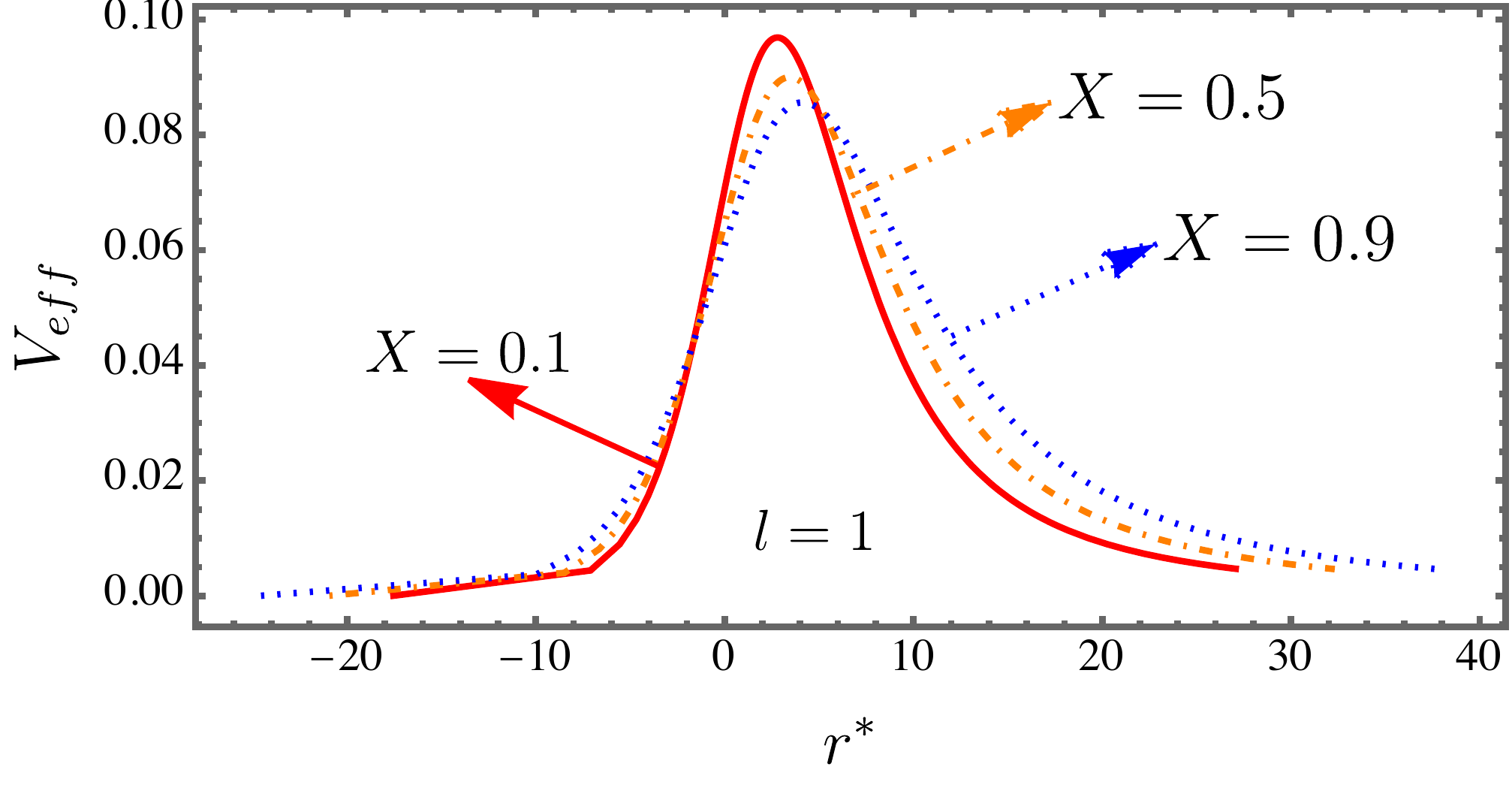}
    \caption{The modifications of $V_{eff}$ caused by different values of $l$ (left) and $X$ (right).}
    \label{effectivepotential}
\end{figure}

In the left panel of Fig. \ref{effectivepotential}, we depict the behavior of the effective potential $V_{\text{eff}}$ as a function of $r^{*}$ for different values of $l$ while maintaining a fixed value of $X$ ($X=0.5$). On the right panel of Fig. \ref{effectivepotential}, we present the behavior for a constant value of $l$ ($l=1$) while varying the parameter $X$ across different values.

 The figure also showcases the modifications induced by parameter $X$, which triggers the Lorentz violation. It is worth noting that the tortoise coordinate $r^{*}$ is a transcendental function of $r$, and making a ``parametric plot'' is the only way to visualize $V_{eff}$ against $r^{*}$. As it is expected, if we consider the limit where $X \rightarrow 0$, we recover the simplest spherically symmetric black hole, i.e., the Schwarzschild case.


\subsection{The WKB method}

In this section, our objective is to derive stationary solutions, characterized by the assumption that $\Psi(t,r)$ can be written as $\Psi(t,r) = e^{-i\omega t} \psi(r)$, where $\omega$ represents the frequency. By adopting this assumption, we can effectively separate the time--independent component of Eq. (\ref{schordingereq}) in the following manner:
\ie
\frac{\partial^{2} \psi}{\partial r^{*2}} - \left[  \omega^{2} - V_{eff}(r^{*})\right]\psi = 0.\label{timeindependent}
\fe
In order to solve Eq. (\ref{timeindependent}), it is crucial to take into account the appropriate boundary conditions. In our particular scenario, the solutions that satisfy the desired conditions are those which exhibit pure ingoing behavior near the horizon. This concept is illustrated below:
\begin{equation}\label{boundaryconditions11}
    \psi^{\text{in}}(r^{*}) \sim 
\begin{cases}
	C_{l}(\omega) e^{-i\omega r^{*}} & ( r^{*}\rightarrow - \infty)\\
	A^{(-)}_{l}(\omega) e^{-i\omega r^{*}} + A^{(+)}_{l}(\omega) e^{+i\omega r^{*}} & (r^{*}\rightarrow + \infty),
\end{cases}
\end{equation}
The complex constants $C_{l}(\omega)$, $A^{(-)}_{l}(\omega)$, and $A^{(+)}_{l}(\omega)$ play a significant role in the following discussion. In general, the \textit{quasinormal} modes of a black hole can be defined as the frequencies ${\omega_{nl}}$ for which $A^{(-)}_{l}(\omega_{nl})=0$. This condition ensures that the modes correspond to a purely outgoing wave at spatial infinity and a purely ingoing wave at the event horizon. Here, the integers $n$ and $l$ represent the overtone and multipole numbers, respectively. The spectrum of \textit{quasinormal} modes is determined by the eigenvalues of Eq. (\ref{timeindependent}). To analyze these modes, we employ the WKB method, i.e., a semi--analytical approach that capitalizes on the analogy with quantum mechanics.

The application of the WKB approximation to calculate the \textit{quasinormal} modes in the context of particle scattering around black holes was initially introduced by Schutz and Will \cite{schutz1985black}. Subsequently, Konoplya made further advancements in this technique \cite{konoplya2003quasinormal,konoplya2004quasinormal}. However, it is important to note in order to make this method valid, the potential must exhibit a barrier--like shape, approaching constant values as $r^{*} \rightarrow \pm \infty$. By fitting the power series of the solution near the turning points of the maximum potential, the \textit{quasinormal} modes can be obtained \cite{santos2016quasinormal}. The Konoplya formula for calculating them is expressed as follows:
\ie \label{boundary}
\frac{i(\omega^{2}_{n}-V_{0})}{\sqrt{-2 V^{''}_{0}}} - \sum^{6}_{j=2} \Lambda_{j} = n + \frac{1}{2}.
\fe
In the aforementioned expression, Konoplya's formula for the \textit{quasinormal} modes incorporates several key elements. The term $V^{''}_{0}$ represents the second derivative of the potential evaluated at its maximum point $r_{0}$, while $\Lambda_{j}$  constants that depend on the effective potential and its derivatives at the maximum. Notably, it is important to mention that a recent development in the field introduced a 13th--order WKB approximation, as proposed by Matyjasek and Opala in the literature \cite{matyjasek2017quasinormal}.

Tables \ref{demo-table1}, \ref{demo-table2}, and \ref{demo-table3} present the \textit{quasinormal} frequencies calculated using the sixth--order WKB method for different values of the parameter $X$. The tables are specifically divided based on the multipole number $l$, with Table \ref{demo-table1} corresponding to $l=0$, Table \ref{demo-table2} to $l=1$, and Table \ref{demo-table3} to $l=2$. 

Importantly, it is noteworthy that the modes associated with the scalar field exhibit negative values in their imaginary part. This indicates that these modes decay exponentially over time, representing the dissipation of energy through scalar waves. These findings align with previous studies investigating scalar, electromagnetic, and gravitational perturbations in spherically symmetric geometries \cite{berti2009quasinormal,konoplya2011quasinormal,heidari2023gravitational,chen2023quasinormal}.

In our investigation, we have observed a decrease in the absolute values of imaginary parts of the \textit{quasinormal} modes as Lorentz--violating parameter increases in general.
The parameter $X$ plays a critical role in this scenario as it governs the damping of the scalar waves. Depending on the its value, the damping can occur at a faster or slower rate.

\begin{figure}
    \centering
    \includegraphics[scale=0.446]{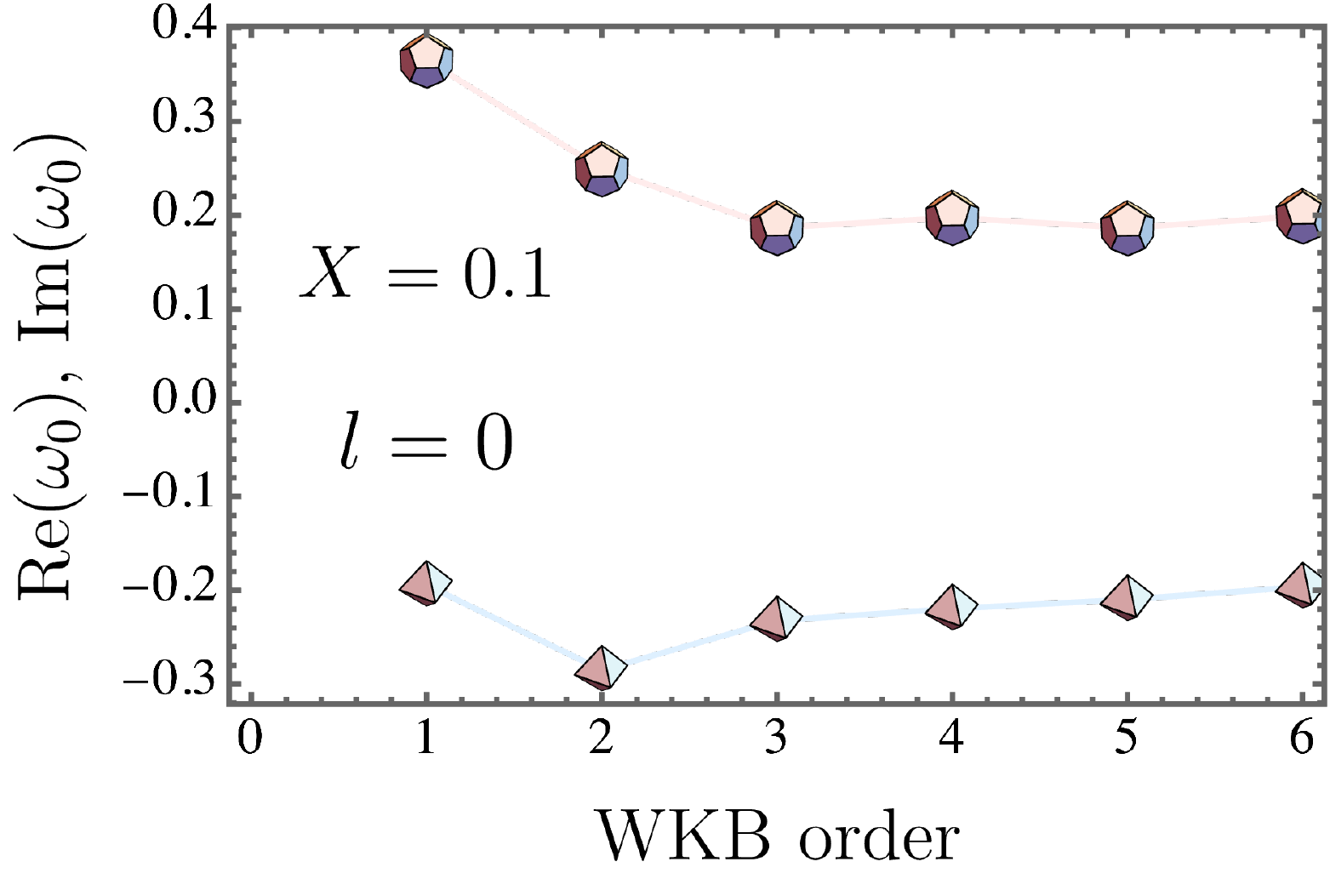}
    \includegraphics[scale=0.45]{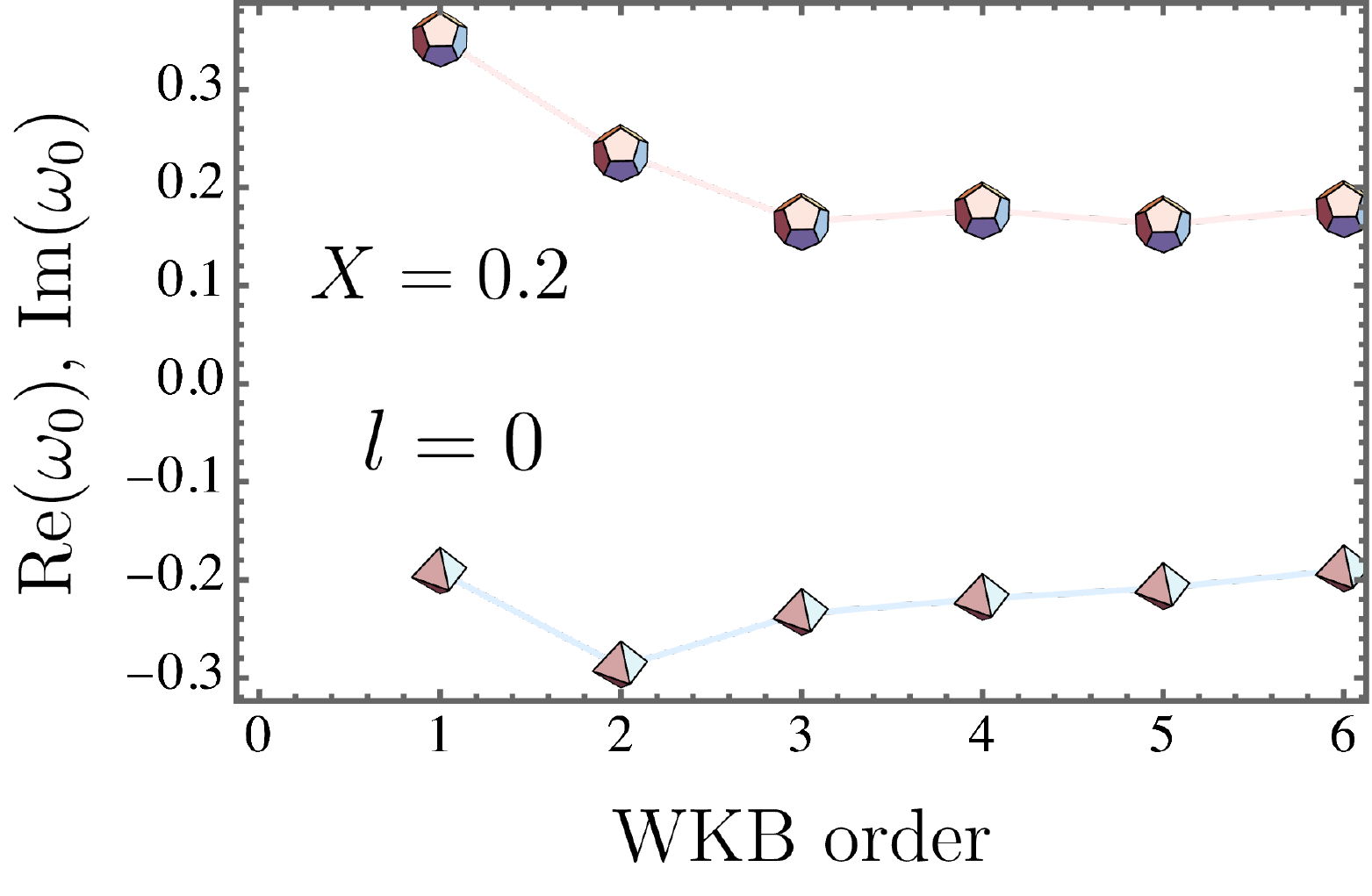}
    \includegraphics[scale=0.45]{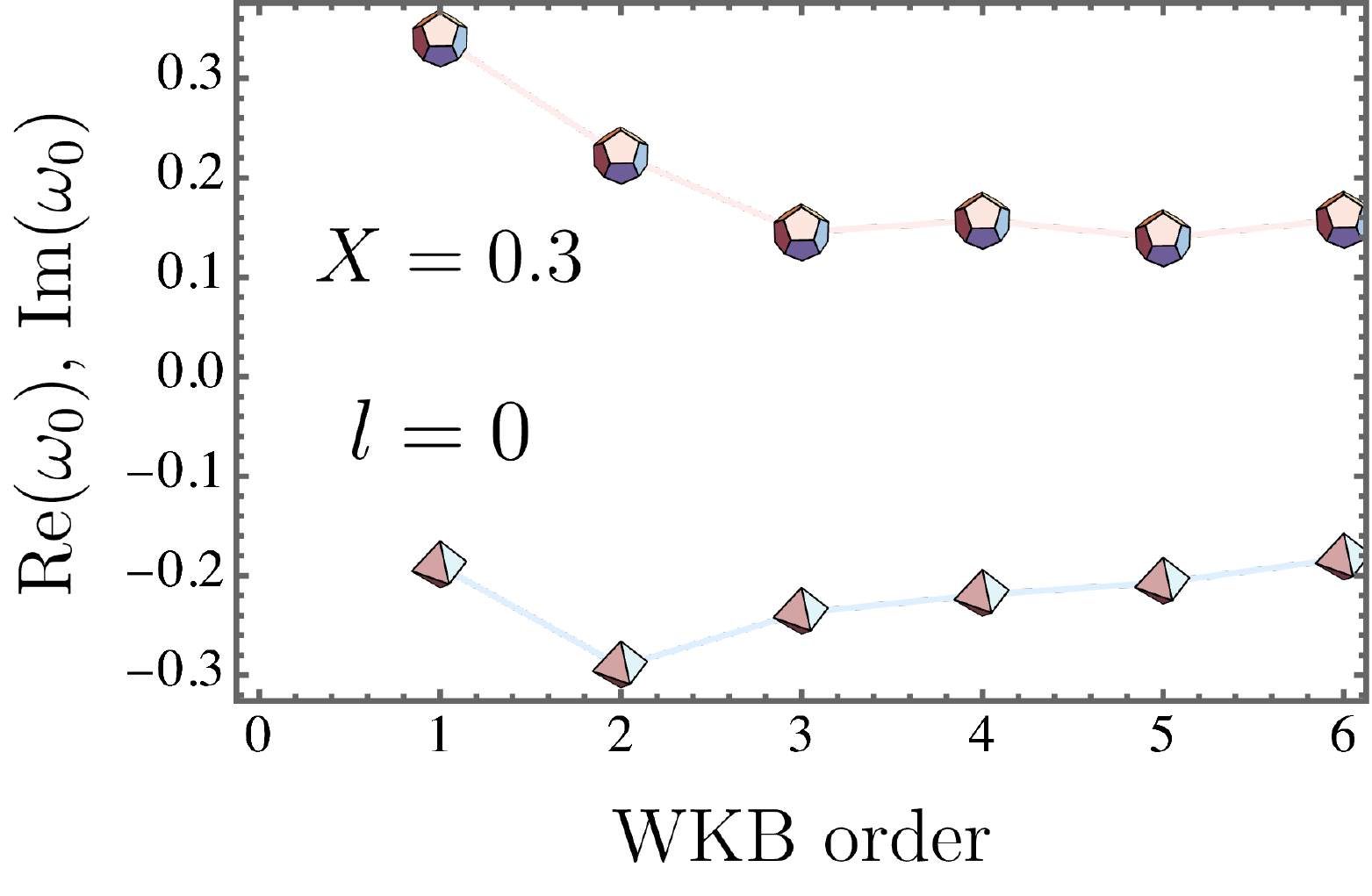}
    \includegraphics[scale=0.45]{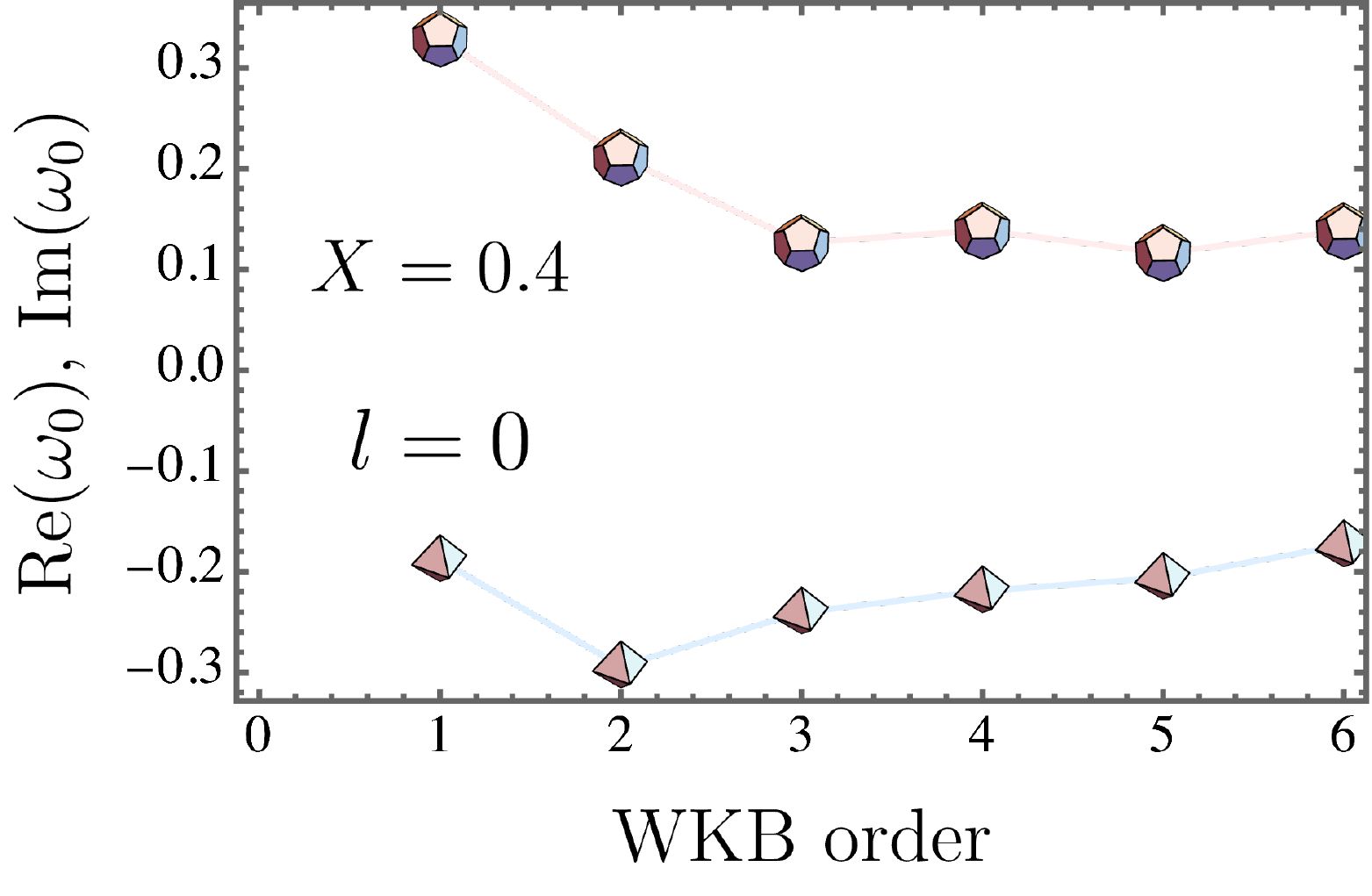}
    \caption{Real (top lines) and imaginary (bottom lines) parts of the \textit{quasinormal} modes to different WKB orders, considering different values of $X$ (for fixed $l=0$).}
    \label{wkborder1}
\end{figure}

In Fig. \ref{wkborder1}, we present a plot showcasing the real and imaginary parts of the \textit{quasinormal} frequencies as a function of the WKB order. Throughout the plot, we maintain the multipole and overtone numbers fixed at $l=0$ and $n=0$, respectively. It is evident that in all cases, the real and imaginary parts of $\omega_{0}$ exhibit convergence as the WKB order increases. This convergence serves as an indication of the effectiveness of such an approach in providing accurate approximations for the \textit{quasinormal} modes.

Nevertheless, it is crucial to acknowledge that increasing the number of WKB orders does not guarantee convergence in general, as the WKB series converges asymptotically. Thus, while the method proves reliable in this particular scenario, caution should be exercised when applying it to other cases.

\begin{table}[!h]
\begin{center}
\begin{tabular}{c c c c} 
 \hline\hline
 $X$ & $\omega_{0}$ & $\omega_{1}$ & $\omega_{2}$ \\ [0.2ex] 
 \hline 
 0.1 & 0.1991 - 0.1963$i$ & 0.1554 - 0.6856$i$ & 0.4191 - 0.8840$i$ \\ 
 
 0.2 & 0.1782 - 0.1900$i$ & 0.1339 - 0.6811$i$ & 0.4743 - 0.8021$i$ \\
 
 0.3 & 0.1581 - 0.1826$i$ & 0.1134 - 0.6757$i$ & 0.5555 - 0.7121$i$ \\
 
 0.4 & 0.1381 - 0.1737$i$ & 0.0935 - 0.6695$i$ & 0.6678 - 0.6229$i$ \\
 
 0.5 & 0.1180 - 0.1628$i$ & 0.0740 - 0.6614$i$ & 0.8115 - 0.5445$i$ \\
 
 0.6 & 0.9754 - 0.1494$i$ & 0.0547 - 0.6519$i$ & 0.9754 - 0.4856$i$ \\
 
 0.7 & 0.0738 - 0.1326$i$ & 0.0353 - 0.6411$i$ & 1.1491 - 0.4453$i$ \\
 
 0.8 & 0.0457 - 0.1108$i$ & 0.0154 - 0.6282$i$ & 1.3302 - 0.4185$i$ \\
 
 0.9 & 0.0037 - 0.0835$i$ & 0.0050 - 0.6145$i$ & 1.5189 - 0.4010$i$ \\ [0.2ex] 
 \hline \hline
\end{tabular}
\caption{\label{demo-table1}The \textit{quasinormal} frequencies by using sixth--order WKB approximation for different values of Lorentz--violating parameter $X$. In this case, the multipole number is $l=0$.}
\end{center}
\end{table}

\begin{table}[!h]
\begin{center}
\begin{tabular}{c c c c} 
 \hline\hline
 $X$ & $\omega_{0}$ & $\omega_{1}$ & $\omega_{2}$ \\ [0.2ex] 
 \hline 
 0.1 & 0.5719 - 0.1947$i$ & 0.5138 - 0.6118$i$ & 0.4464 - 1.0841$i$ \\ 
 
 0.2 & 0.5596 - 0.1941$i$ & 0.5004 - 0.6106$i$ & 0.4326 - 1.0839$i$ \\
 
 0.3 & 0.5487 - 0.1934$i$ & 0.4884 - 0.6095$i$ & 0.4203 - 1.0836$i$ \\
 
 0.4 & 0.5389 - 0.1928$i$ & 0.4778 - 0.6085$i$ & 0.4093 - 1.0834$i$ \\
 
 0.5 & 0.5303 - 0.1923$i$ & 0.4683 - 0.6075$i$ & 0.3995 - 1.0831$i$ \\
 
 0.6 & 0.5226 - 0.1918$i$ & 0.4598 - 0.6066$i$ & 0.3907 - 1.0828$i$ \\
 
 0.7 & 0.5157 - 0.1913$i$ & 0.4522 - 0.6057$i$ & 0.3828 - 1.0825$i$ \\
 
 0.8 & 0.5096 - 0.1908$i$ & 0.4455 - 0.6048$i$ & 0.3757 - 1.0821$i$ \\
 
 0.9 & 0.5043 - 0.1904$i$ & 0.4395 - 0.6040$i$ & 0.3694 - 1.0817$i$ \\ [0.2ex] 
 \hline \hline
\end{tabular}
\caption{\label{demo-table2}The \textit{quasinormal} frequencies by using sixth--order WKB approximation for different values of Lorentz--violating parameter $X$. In this case, the multipole number is $l=1$.}
\end{center}
\end{table}

\begin{table}[!h]
\begin{center}
\begin{tabular}{c c c c} 
 \hline\hline
 $X$ & $\omega_{0}$ & $\omega_{1}$ & $\omega_{2}$ \\ [0.2ex] 
 \hline 
 0.1 & 0.9517 - 0.1932$i$ & 0.9115 - 0.5906$i$ & 0.8438 - 1.0172$i$ \\ 
 
 0.2 & 0.9380 - 0.1929$i$ & 0.8973 - 0.5901$i$ & 0.8289 - 1.0171$i$ \\
 
 0.3 & 0.9260 - 0.1927$i$ & 0.8848 - 0.5896$i$ & 0.8158 - 1.0169$i$ \\
 
 0.4 & 0.9155 - 0.1925$i$ & 0.8738 - 0.5892$i$ & 0.8043 - 1.0168$i$ \\
 
 0.5 & 0.9062 - 0.1922$i$  & 0.8641 - 0.5887$i$  & 0.7941 - 1.0166$i$ \\
 
 0.6 & 0.8981 - 0.1921$i$ & 0.8556 - 0.5883$i$ & 0.7852 - 1.0164$i$ \\
 
 0.7 & 0.8910 - 0.1919$i$ & 0.8483 - 0.5879$i$ & 0.7775 - 1.0162$i$ \\
 
 0.8 & 0.8849 - 0.1917$i$ & 0.8419 - 0.5876$i$ & 0.7708 - 1.0160$i$ \\
 
 0.9 & 0.8797 - 0.1916$i$ & 0.8365 - 0.5872$i$ & 0.7650 - 1.0157$i$ \\ [0.2ex] 
 \hline \hline
\end{tabular}
\caption{\label{demo-table3}The \textit{quasinormal} frequencies by using sixth--order WKB approximation for different values of Lorentz--violating parameter $X$. In this case, the multipole number is $l=2$.}
\end{center}
\end{table}

An additional crucial aspect to consider is the utilization of the WKB method for studying scattering, which necessitates appropriate boundary conditions. In this context, our objective is to determine the reflection and transmission coefficients, which bear resemblance to those encountered in quantum mechanics for tunneling phenomena. To compute these coefficients, we take advantage of the fact that $(\omega^{2}_{n}-V_{0})$ is purely real and derive the following expression:
\ie
\Upsilon= \frac{i(\Tilde{\omega}^{2}-V_{0})}{\sqrt{-2 V^{''}_{0}}} - \sum_{j=2}^{6} \Lambda_{j}(\Upsilon).
\fe
To determine the reflection and transmission coefficients, we analyze the scattering process using the semi--classical WKB approach, which has been the subject of recent research in the field \cite{konoplya2020quantum,konoplya2019higher,campos2022quasinormal}. These coefficients are associated with the effective potential and are represented by complex functions denoted as $\Lambda_{j}(\Upsilon)$, where $\Upsilon$ is a purely imaginary quantity, and $\Tilde{\omega}$ corresponds to the real frequency associated with the \textit{quasinormal} modes. The expressions for the reflection and transmission coefficients are given as follows:
\ie
|R|^{2} = \frac{|A_{l}^{(+)}|^{2}}{|A_{l}^{(-)}|^{2}} = \frac{1}{1+e^{-2i\pi \Upsilon}},
\fe
\ie
|T|^{2} = \frac{|C_{l}|^{2}}{|A_{l}^{(-)}|^{2}} = \frac{1}{1+e^{+2i\pi \Upsilon}}.
\fe
where, $A_{l}^{(+)}$,$A_{l}^{(-)}$ and $C_{l}$ are the complex numbers which can be found according to the boundary condition in Eq. (\ref{boundaryconditions11}).
\begin{figure}
    \centering
    \includegraphics[scale=0.4]{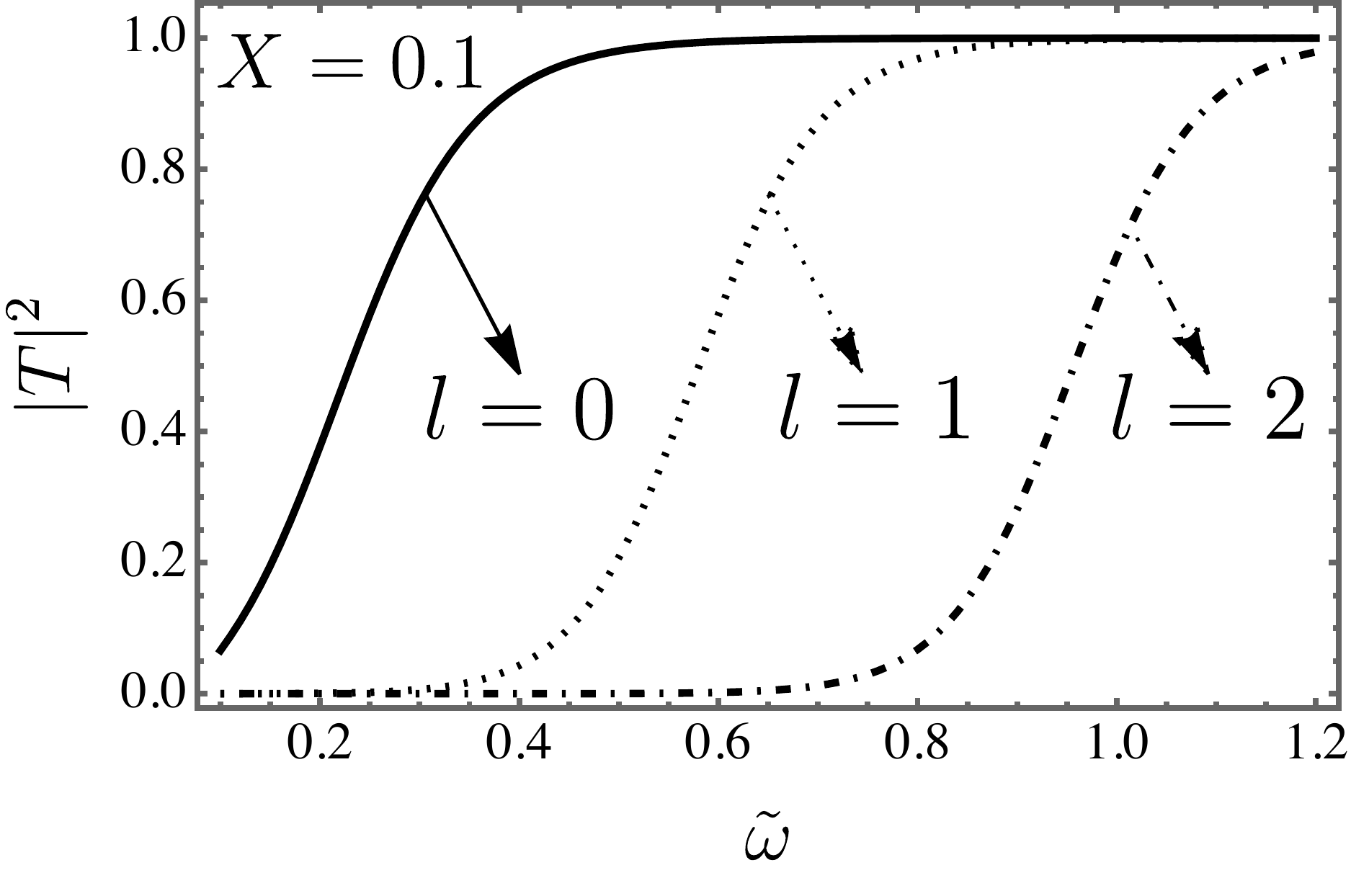}
    \includegraphics[scale=0.4]{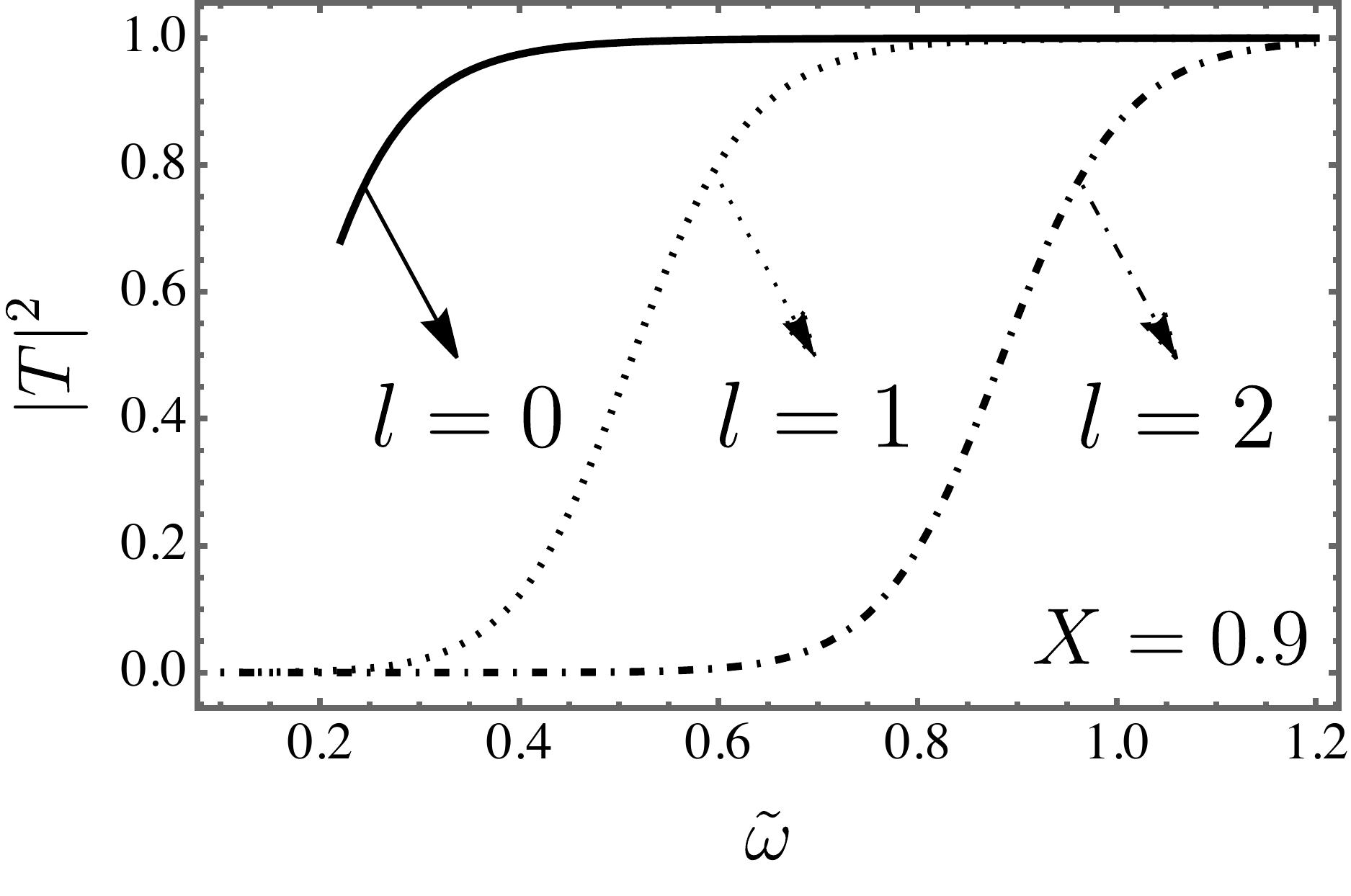}
    \caption{Transmission coefficients for distinct values of $l$ and $X$.}
    \label{transmissioncoefficients}
\end{figure}
\begin{figure}
    \centering
    \includegraphics[scale=0.4]{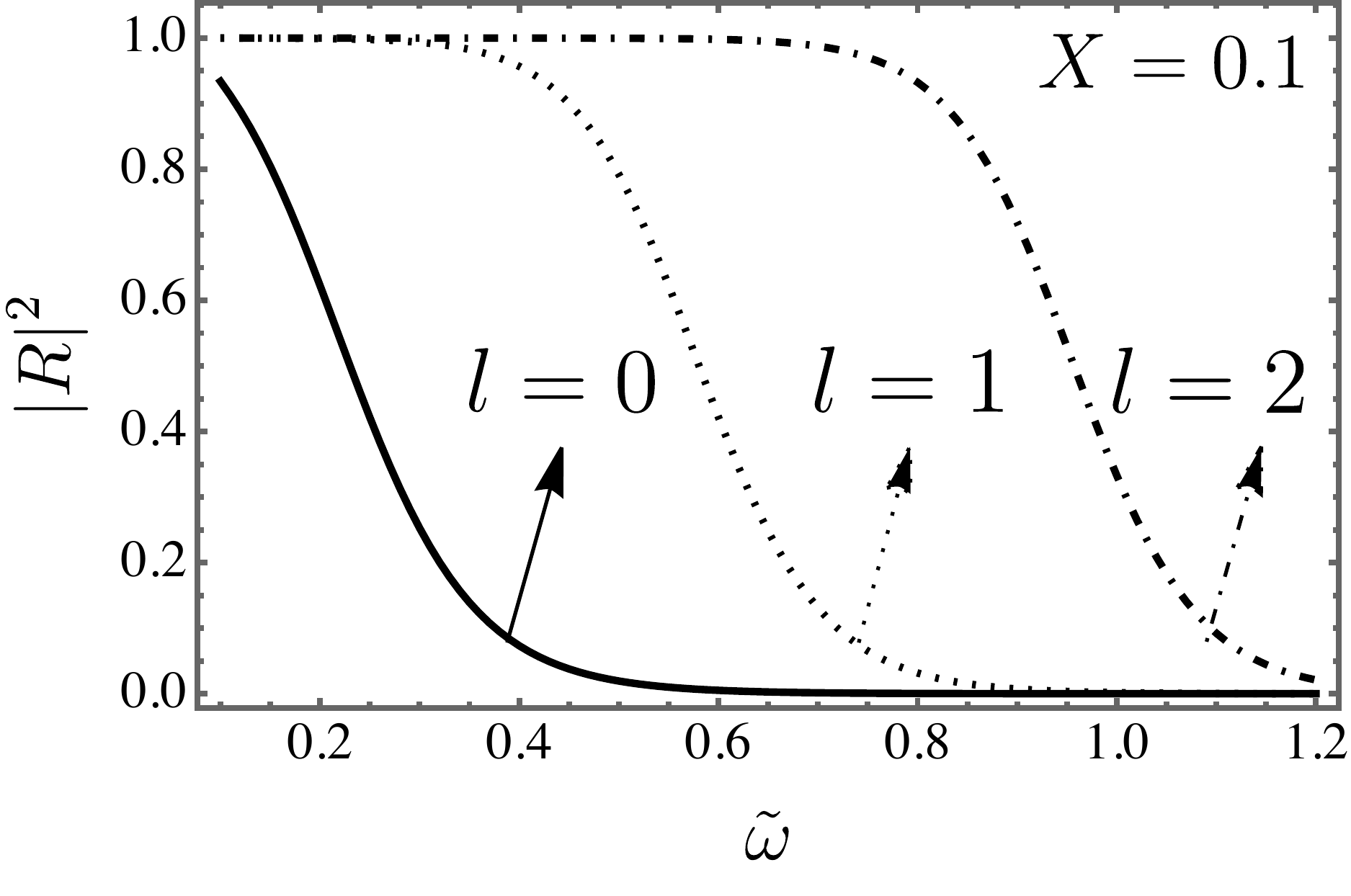}
    \includegraphics[scale=0.4]{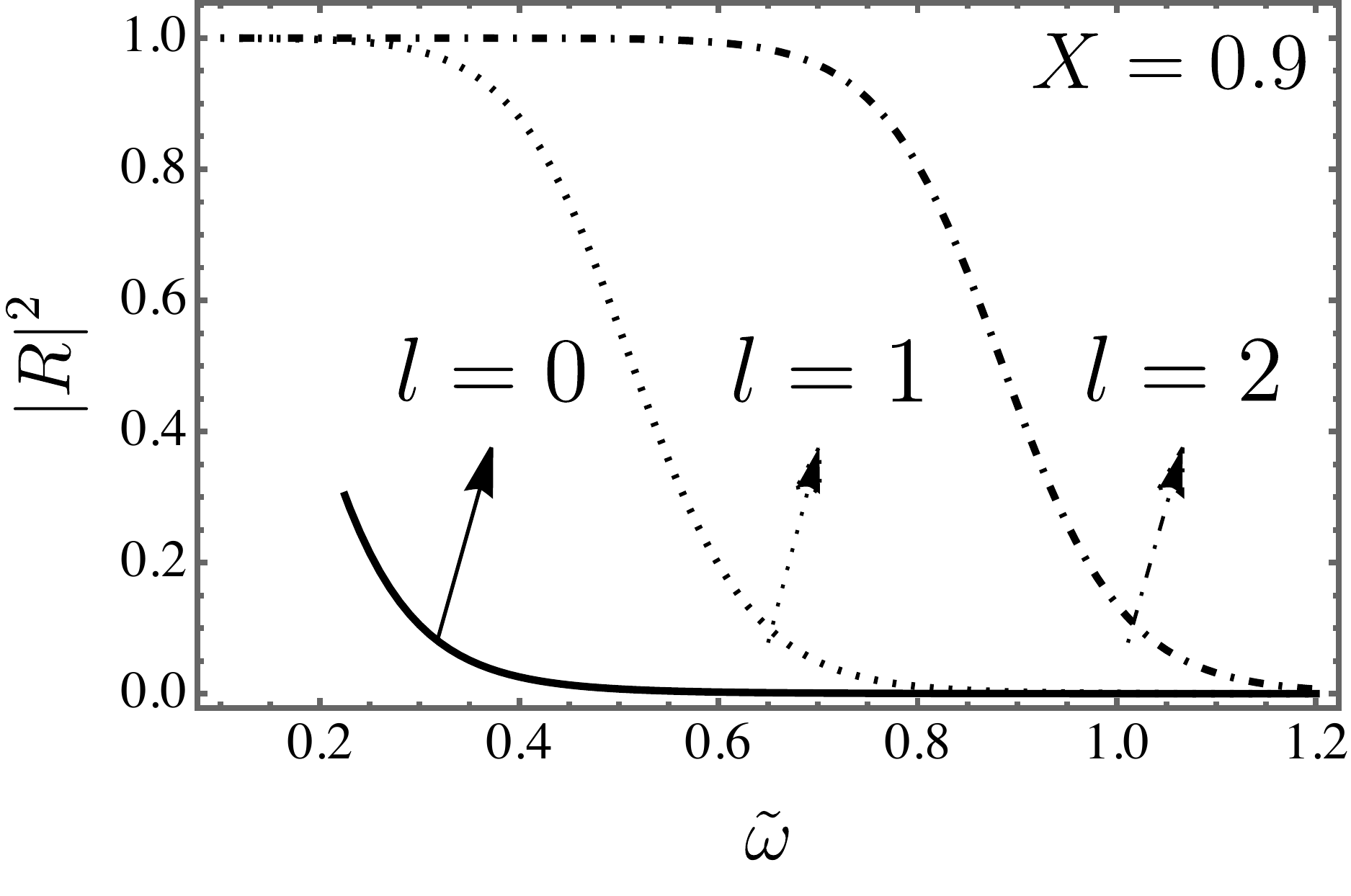}
    \caption{Reflection coefficients for distinct values of $l$ and $X$.}
    \label{reflectionscoefficients}
\end{figure}
In Figs. \ref{transmissioncoefficients} and \ref{reflectionscoefficients}, the transmission and reflection coefficients are displayed, respectively, for various values of the Lorentz--violating parameter and multipole numbers. Notably, it is observed that for a fixed value of $X$, increasing the multipole number $l$ results in a rightward shift of the transmission coefficients. Conversely, when $l$ is kept constant and $X$ is increased, the reflection coefficients shift in the opposite direction. 

However, it is important to highlight that the 6th--order WKB approximation does not yield accurate results for $l=0$ in the context of transmission and reflection coefficients. This observation aligns with findings from previous studies \cite{campos2022quasinormal,konoplya2019higher,heidari2023gravitational}.


\section{Time--Domain solution}

To thoroughly investigate the impact of the \textit{quasinormal} spectrum on time--dependent scattering phenomena, a detailed analysis of scalar perturbations in the time domain is necessary. However, given the intricacy of our effective potential, a more precise approach is required to gain deeper insights. In this regard, we employ the characteristic integration method developed by Gundlach and collaborators \cite{gundlach1994late} as an effective tool to study the problem effectively. By utilizing this approach, we can acquire valuable insights into the role of \textit{quasinormal} modes in time--dependent scattering scenarios, which have significant implications for the investigation of black holes and related phenomena.

The methodology presented in Ref. \cite{gundlach1994late} revolves around the utilization of light--cone coordinates, represented by $u = t - x$ and $v = t + x$. These coordinates enable the reformulation of the wave equation in a more suitable manner, allowing for a comprehensive examination of the system
\ie
\left(  4 \frac{\partial^{2}}{\partial u \partial v} + V(u.v)\right) \Psi (u,v) = 0 \label{timedomain}.
\fe
To achieve efficient integration of the aforementioned expression, a discretization scheme can be employed, making use of a simple finite--difference method and numerical techniques. This approach allows for the effective numerical integration of the equation, providing accurate and efficient results
\ie
\Psi(N) = -\Psi(S) + \Psi(W) + \Psi(E) - \frac{h^{2}}{8}V(S)[\Psi(W) + \Psi(E)] + \mathcal{O}(h^{4}),
\fe
where $S=(u,v)$, $W=(u+h,v)$, $E=(u,v+h)$, and $N=(u+h,v+h)$, where $h$ represents the overall grid scale factor. The null surfaces $u=u_{0}$ and $v=v_{0}$ are of particular significance, as they serve as locations where the initial data are specified. In our investigation, we have chosen to employ a Gaussian profile centered at $v=v_{c}$ with a width of $\sigma$, which is selected on the null surface $u=u_{0}$
\ie
\Psi(u=u_{0},v) = A e^{-(v-v_{*})^{2}}/2\sigma^{2}, \,\,\,\,\,\, \Psi(u,v_{0}) = \Psi_{0}.
\fe
At $v=v_{0}$, a constant initial condition $\Psi(u,v_{0}) = \Psi_{0}$ was imposed, and without loss of generality, we assume $\Psi_{0} = 0$. The integration process then proceeds along the $u=$const. lines in the direction of increasing $v$, once the null data are specified. 

Furthermore, the results of our investigation into the scalar test field are presented in this study. For the sake of convenience, we set $m = 1$, and the null data were defined by a Gaussian profile centered at $u = 10$ on the $u=0$ surface, with a width of $\sigma = 3$, and $\Psi_{0}=0$. The grid was established to cover the ranges $u \in [0,200]$ and $v \in [0,200]$, with grid points sampled to yield an overall grid factor of $h = 0.1$. 

\begin{figure}
    \centering
    \includegraphics[scale=0.45]{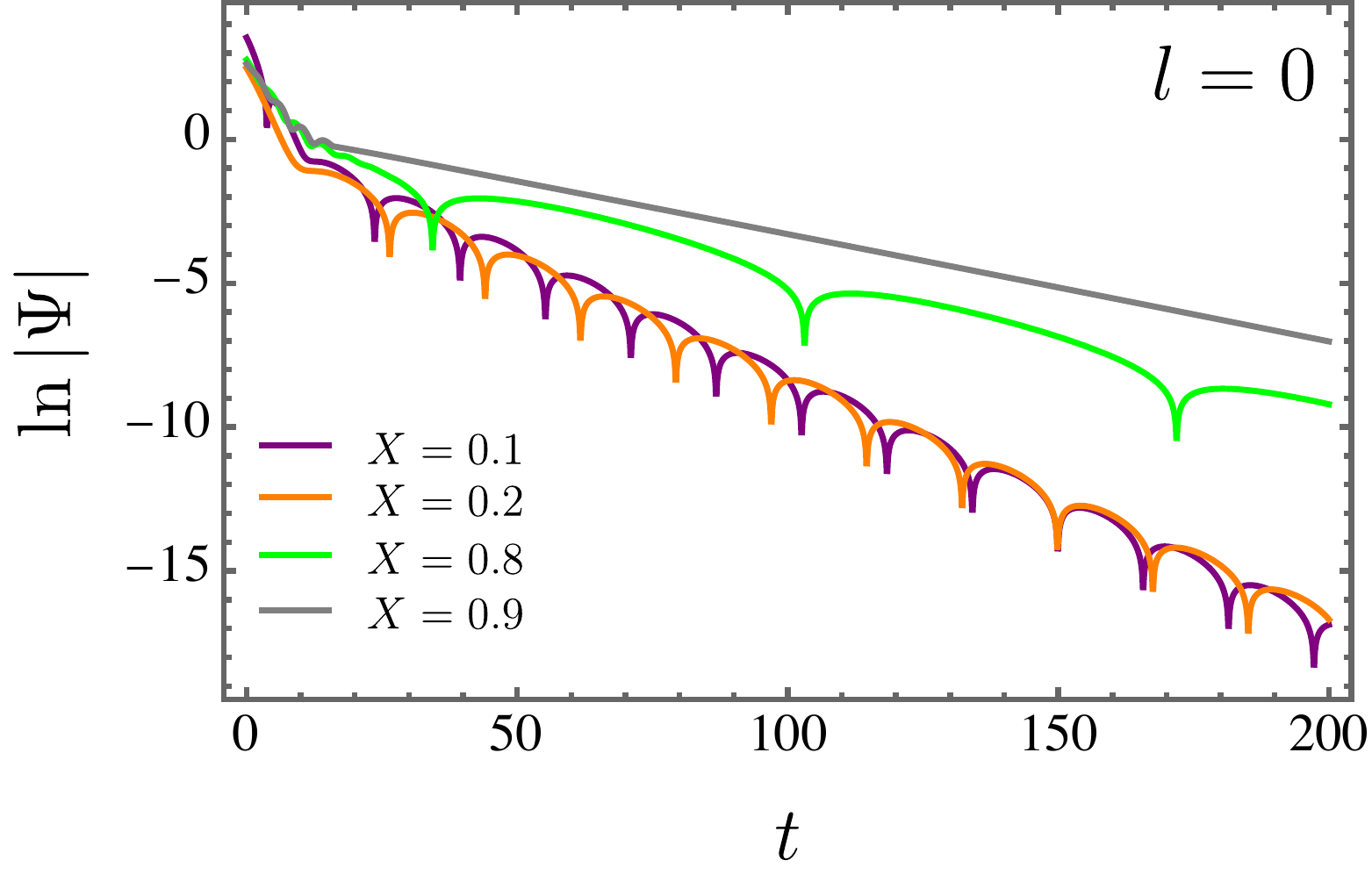}
    \includegraphics[scale=0.45]{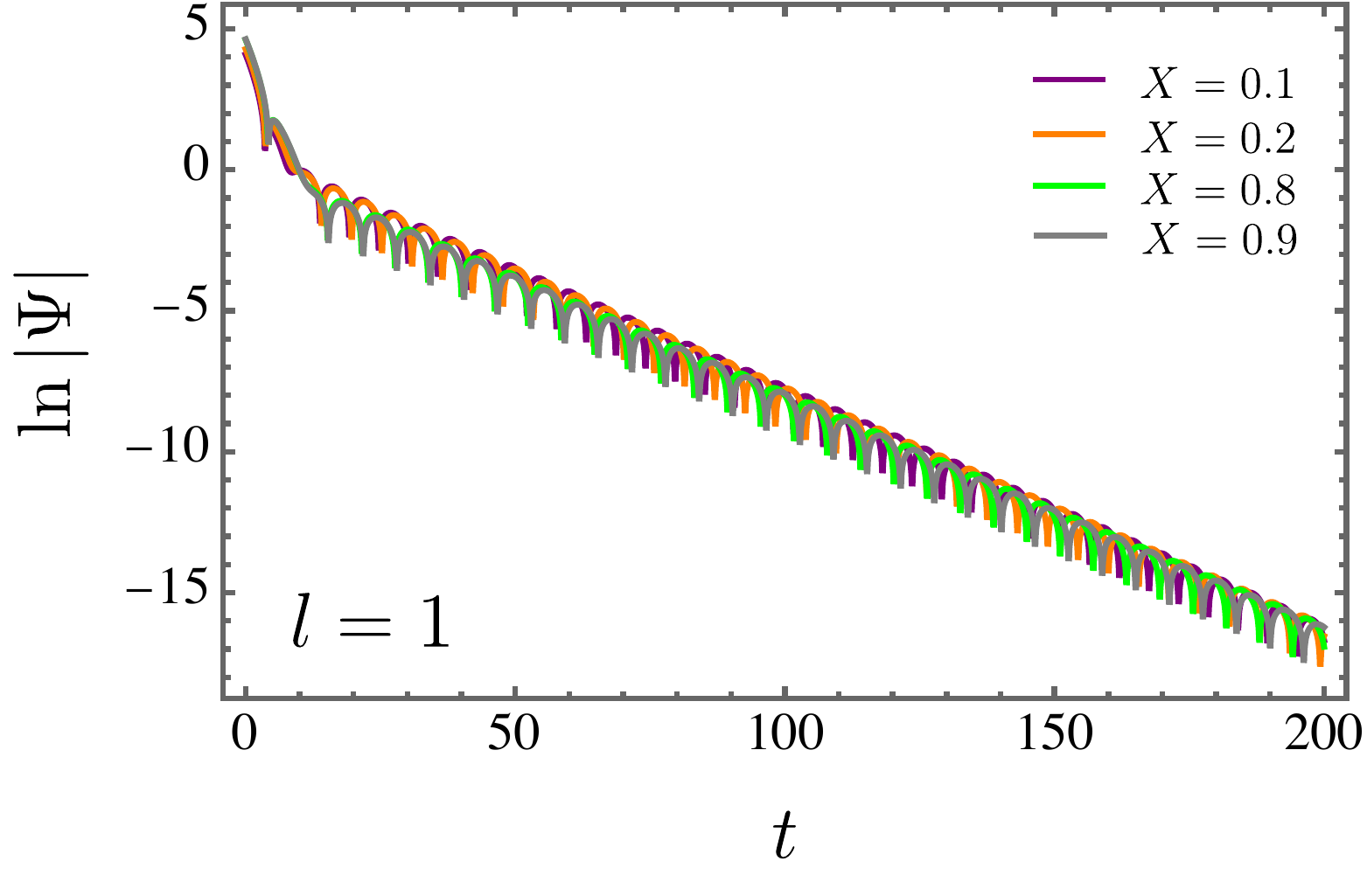}
    \includegraphics[scale=0.45]{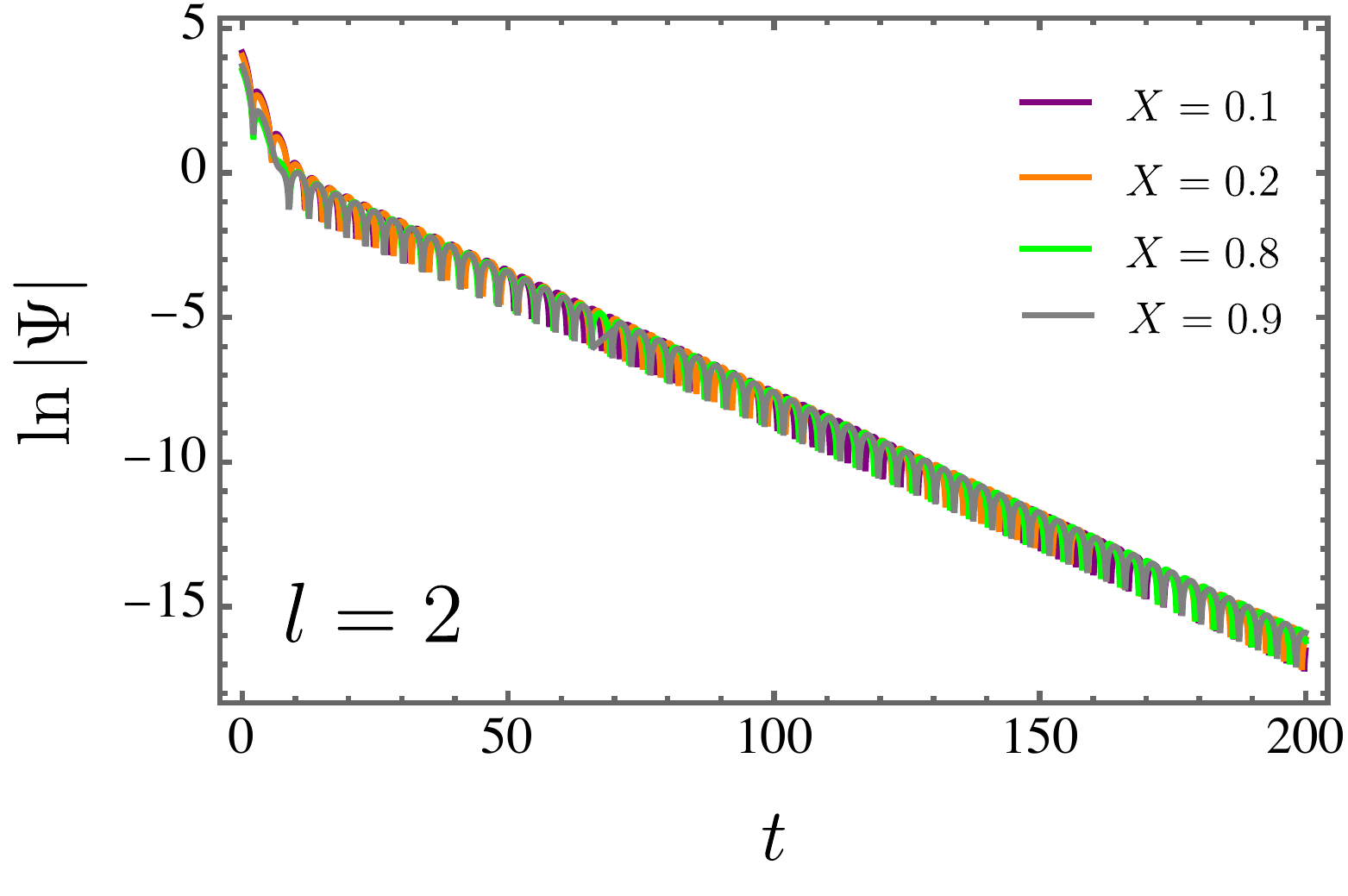}
    \caption{Time domain profiles of scalar perturbations at $r^{*} = 10M$ are presented for various values of the Lorentz--violating parameter $X$ and $l$.}
    \label{Psiln}
\end{figure}

To validate our findings, we present Fig. \ref{Psiln}, which showcases the typical evolution profiles for different combinations of $X$ and $l$, allowing for a visual comparison and analysis of the results.


\section{Shadows}

To investigate the effects of the Lorentz violation of Eq. (\ref{metric3}) in the shadows, we consider
\begin{equation}\label{action1}
\frac{{\partial \mathcal{S}}}{{\partial \tau }} =  - \frac{1}{2}{g^{\mu \nu }}\frac{{\partial S}}{{\partial {\tau ^\mu }}}\frac{{\partial S}}{{\partial {\tau ^\nu }}}
\end{equation}
where $\mathcal{S}$ is the Jacobi action and $\tau$ is the arbitrary affine parameter. Also, above expression can be decomposed as
\begin{equation}\label{action2}
\mathcal{S} = \frac{1}{2}{m^2}\tau  - Et + L\phi  + {S_r}(r) + {S_\theta }(\theta ).
\end{equation}
We can mathematically represent $S_r(r)$ and $S_{\theta}(\theta)$ as functions that depend on the variables $r$ and $\theta$, respectively. Since we are interested in the photon behavior, as a consequence, the energy $E$ and angular momentum $L$ become constants of motion. By utilizing Eqs. (\ref{action1}) and (\ref{action2}), we can derive the equations governing the path of the photon, known as null geodesic equations
\begin{equation}\label{time}
\frac{{\mathrm{d}t}}{{\mathrm{d}\tau }} = \sqrt{\left(\frac{3 X}{4}+1\right) \left(1-\frac{X}{4}\right)}\frac{E}{f(r)},
\end{equation}
\begin{equation}\label{position}
\frac{{\mathrm{d}r}}{{\mathrm{d}\tau }} = \frac{{\sqrt {\mathcal{R}(r)} }}{{{r^2}}},
\end{equation}
\begin{equation}\label{thetadot}
\frac{{\mathrm{d}\theta }}{{\mathrm{d}\tau }} =\pm \frac{{\sqrt {\mathcal{Q} (r)} }}{{{r^2}}},
\end{equation}
\begin{equation}\label{phidot}
\frac{{\mathrm{d}\varphi }}{{\mathrm{d}\tau }} = \frac{{L\,{{\csc }^2}\theta }}{{{r^2}}},
\end{equation}
where $\mathcal{R}(r)$ and $\mathcal{Q}(\theta)$ defined as 
\begin{equation}
\mathcal{R}(r) =\left(1-\frac{X}{4}\right)^2 {E^2}{r^4} - (\mathcal{K} + {L^2}){r^2}f(r)\sqrt{\frac{\left(1-\frac{X}{4}\right)^3}{\frac{3 X}{4}+1}}
\end{equation}

\begin{equation}
\mathcal{Q} (\theta ) = \mathcal{K} - {L^2}{\cot ^2}\theta.
\end{equation}
The symbol $\mathcal{K}$ represents the Carter constant \cite{carter1968global}. In Eq. (\ref{thetadot}), the plus and minus signs correspond to the photon's motion in the outgoing and ingoing radial directions, respectively. For simplicity and without loss of generality, we fix the angle $\theta$ to $\pi/2$ and restrict our analysis to the equatorial plane. With this simplification, our attention turns to the radial equation, where we introduce the concept of an effective potential $\mathcal{V}_{eff}(r)$
\begin{equation}\label{veff2}
{\left(\frac{{\mathrm{d}r}}{{\mathrm{d}\tau }}\right)^2} + {\mathcal{V}_{eff}}(r) = 0.
\end{equation}
Here, it is defined 
\begin{equation}\label{veffr}
\mathcal{V}_{eff}(r) = (L^2+\mathcal{K})\sqrt{\frac{\left(1-\frac{X}{4}\right)^3}{\frac{3 X}{4}+1}}\frac{f(r)}{{{r^2}}} -\left(1-\frac{X}{4}\right)^2 {E^2},
\end{equation}
and we introduce two parameters in order to better accomplish our analysis
\begin{equation}\label{par}
\xi  = \frac{L}{E}
\text{ and }
\eta  = \frac{\mathcal{K}}{{E^2}}.
\end{equation}
The so--called critical radius, $r_{c}$ (photon sphere), can be found by considering 
\begin{equation}
\left.\mathcal{V}_{eff}\right|_{{r=r_{c}}}=\left.\frac{{\mathrm{d}{\mathcal{V}_{eff}}}}{{\mathrm{d}r}}\right|_{{r=r_{c}}}=0.
\end{equation}
In addition, 
by using Eqs. (\ref{veffr}) and (\ref{par}) and considering $\left.\mathcal{V}_{eff}\right|_{{r=r_{c}}}=0$, we have
\begin{equation}
{\xi ^2} + \eta  =\sqrt{\left(1+\frac{3X}{4}\right)\left(1-\frac{X}{4}\right)} \frac{{r_{c}^2}}{{f({r_{c}})}}.
\end{equation}

To determine the shadow radius, we will utilize the celestial coordinates $\alpha$ and $\beta$ \cite{singh2018shadow}, which are connected to the constants of motion as follows: $\alpha=-\xi$ and $\beta=\pm\sqrt{\eta}$. With these coordinates, the shadow radius can be expressed as:
\begin{equation}
\mathcal{R}_{\text{Shadow}} =\sqrt{{\xi ^2} + \eta}=\sqrt{\alpha^{2}+\beta^{2}}= \frac{{r_{c}}}{{\sqrt{f({r_{c}})}}}  \left(\left(1+\frac{3X}{4}\right)\left(1-\frac{X}{4}\right)\right)^{1/4}.
\end{equation}
In this manner, 
by considering $\left.\frac{{\mathrm{d}{\mathcal{V}_{eff}}}}{{\mathrm{d}r}}\right|_{{r=r_{c}}}=0$, 
and the using the fact that $f(r_{c}) = 1-\frac{2M}{r_{c}}$
one can obtain $r_{c}=3M$. Therefore,
the equation for the shadow radius reads explicitly 
\begin{equation} \label{Rsh}
\mathcal{R}_{\text{Shadow}} = 3\sqrt{3} M  \sqrt[4]{\left(1+\frac{3X}{4}\right)\left(1-\frac{X}{4}\right)}.
\end{equation}

It is obvious that the Eq. (\ref{Rsh}) recovers the shadow radius for ordinary Schwarzschild black hole by considering $X=0$. Based on the EHT horizonscale image of $Sgr A^*$, two constraints have been proposed for shadow radius \cite{vagnozzi2022horizon}, as $		4.55 < {R_{sh}} < 5.22 $ and $ 4.21 < {R_{sh}} < 5.56 $ considered as $1\sigma$ and $2\sigma$, respectively.

\begin{figure}[h]
		\centering
		\includegraphics[scale = 0.5]{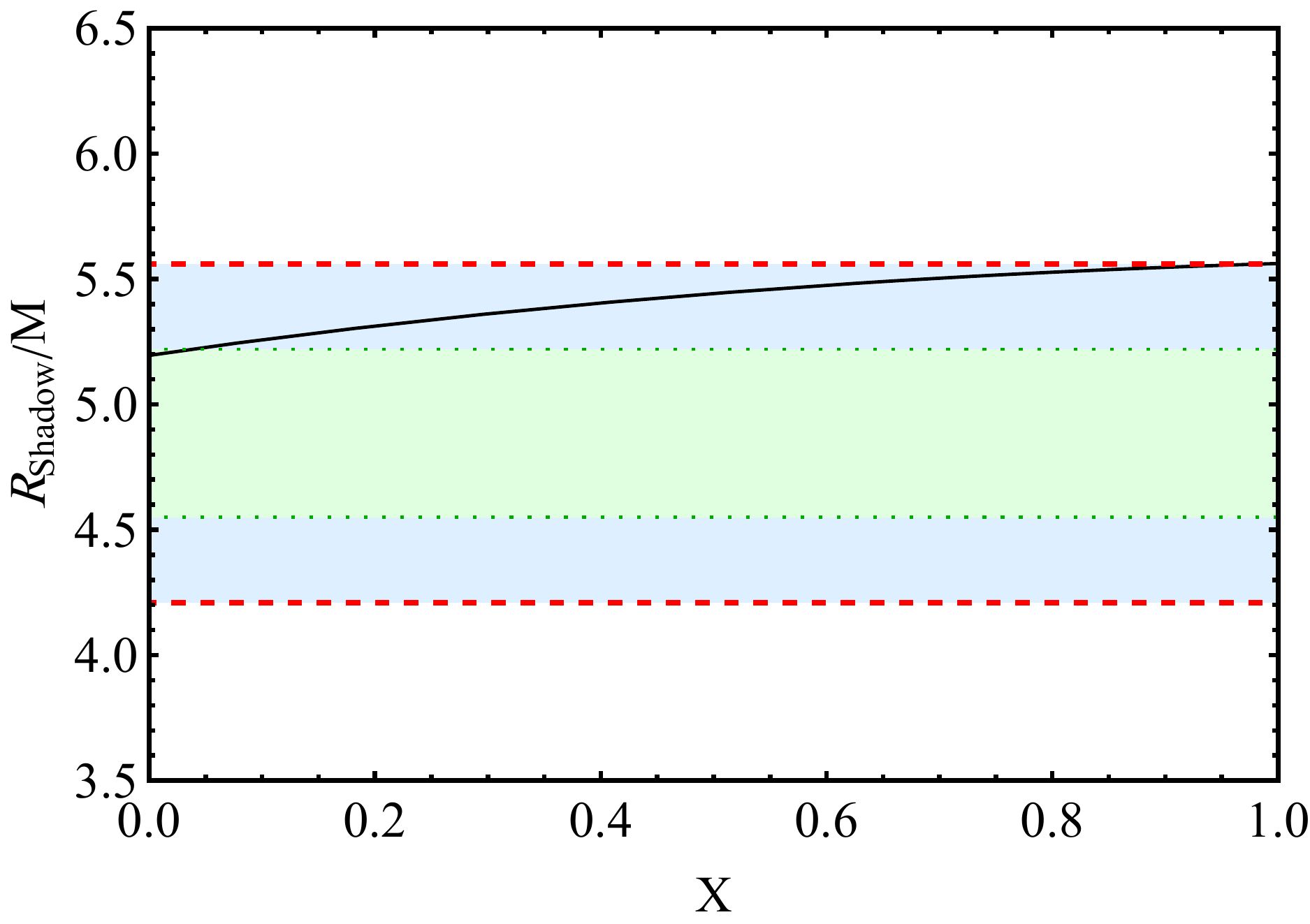}
		\caption{ Shadow radius versus $X$ for $M=1$, the dashed lines represente the experimental constraints of $4.21$, $5.56$ and $4.55$, $5.22$}
		\label{fig:cons}
	\end{figure}

Therefore, we have examined these conditions to explore the limits on the $X$ value. 	For better visualization the shadow radius versus $X$ is plotted for $M = 1$ and the experimental constraints are shown by two pairs of horizontal lines in Fig. \ref{fig:cons}.
	
As depicted in Fig. \ref{fig:cons}, the constraint derived for the parameter $X$ falls within the range of $0<X<0.037$ at the $1\sigma$ confidence level and extends to $0<X<0.86$ at the $2\sigma$ confidence level.

\begin{figure}[!]
	\centering
	\includegraphics[width=80mm]{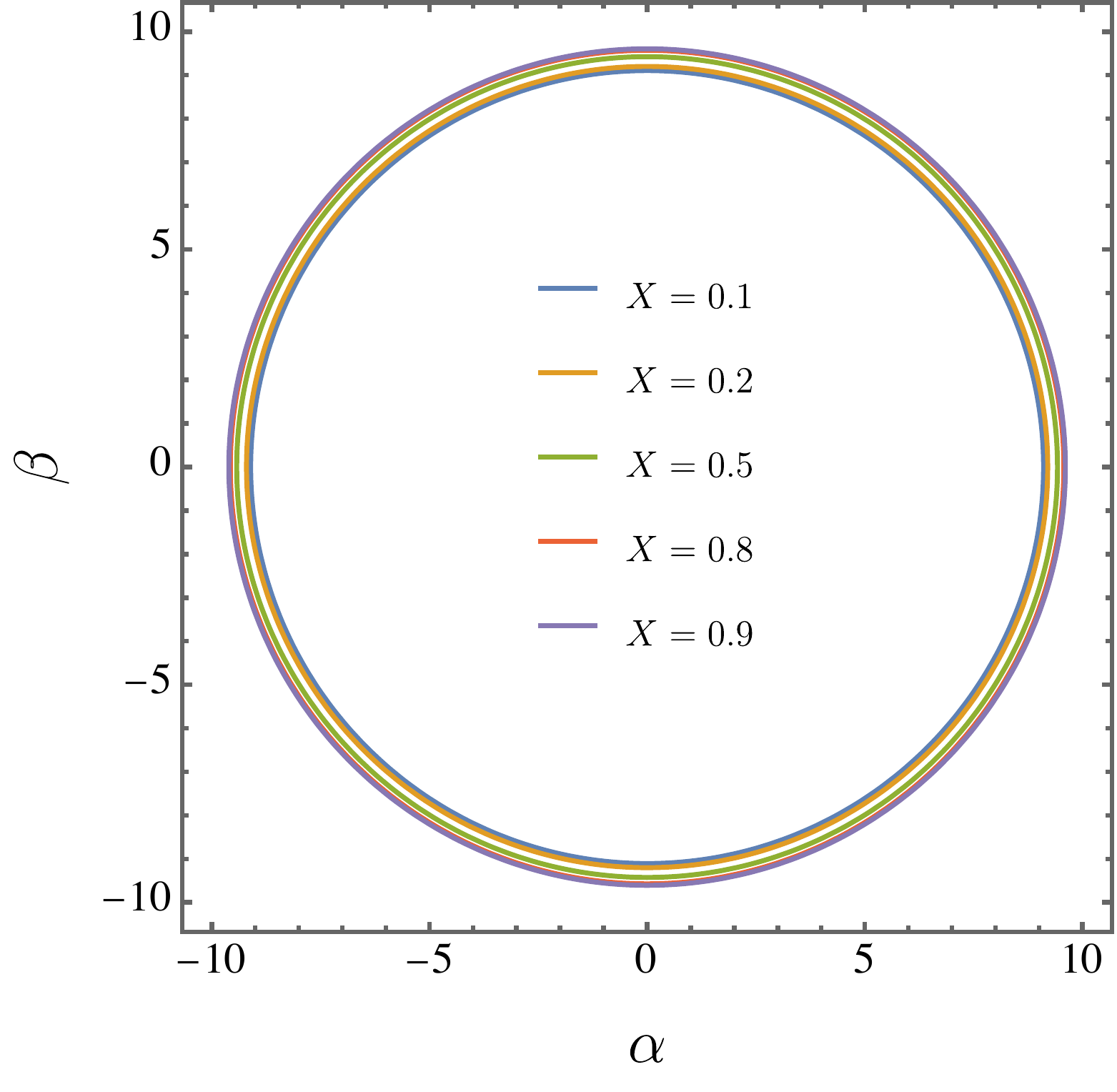}
	\hfill
	\includegraphics[width=80mm]{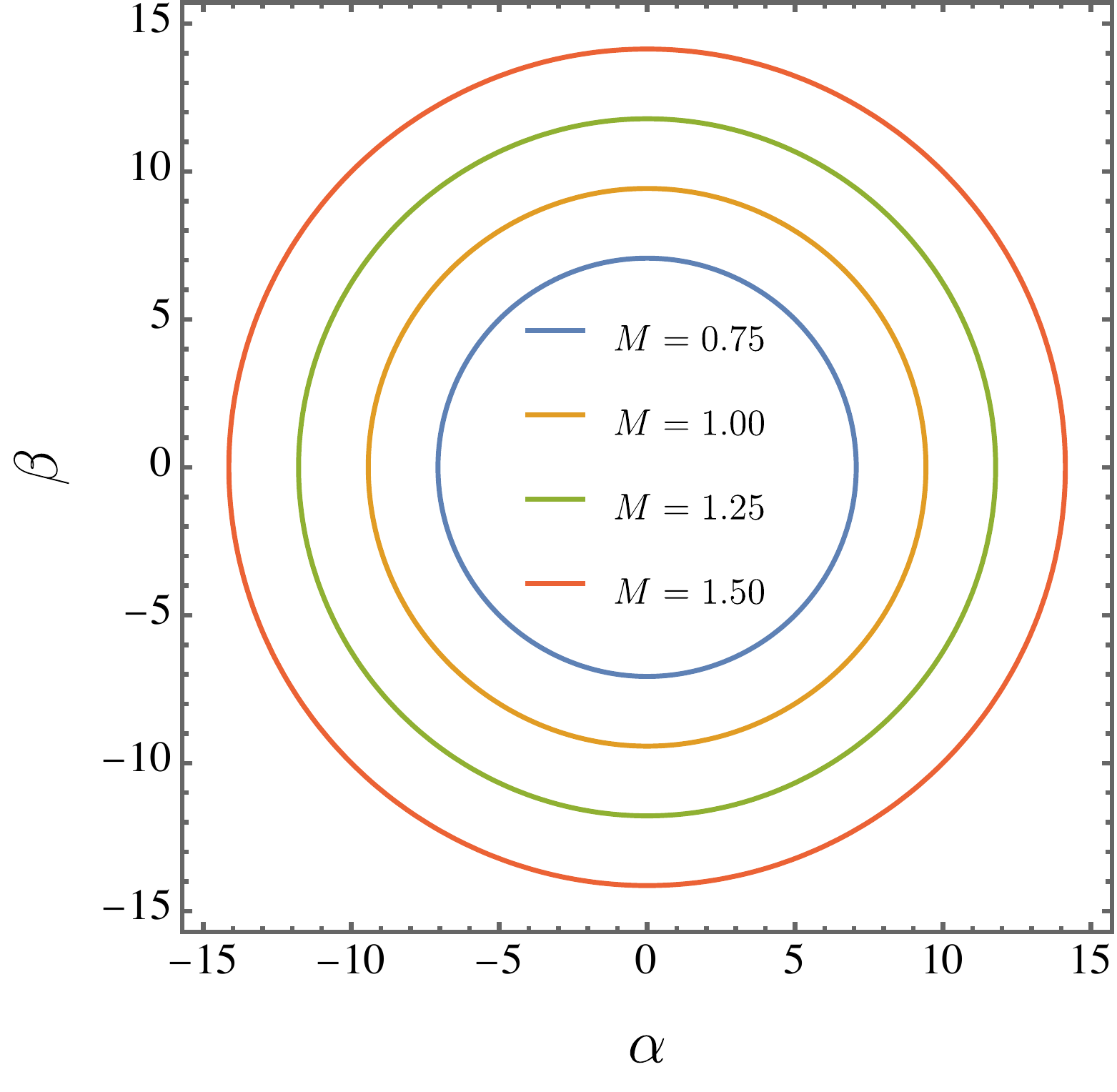}
	\caption{(left) Shadows with mass $M=1$ for diverse values of parameter $X = 0.1, 0.2, 0.5, 0.8, 0.9$. On the other hand, (right) shadows with $X =0.5$, considering a variety of mass: $M=0.75,1.00,1.25,1.50$.}
	\label{fig:shadow}
\end{figure}

Then, in Fig. \ref{fig:shadow}, we display an analysis of the shadows of our black hole for a range of $X$ values on the left side. Remarkably, the shadow radius demonstrates a noticeable augmentation as the parameter $X$ increases. On the right--hand side, we investigate the influence of mass on the same Lorentz--violating parameter, namely, $X=0.5$. As illustrated in the figure, the shadow radius experiences a notable enlargement when the initial mass $M$ transitions from $0.75$ to $1.50$. In addition, since we are dealing with a real positive defined values of radius, $\mathcal{R}_{\text{Shadow}}$ turns out to be bounded from above, i.e., having its maximum value for $X=0.86$.

\section{Time delay and Deflection angle}

The investigation of time delay in Lorentz--violating scenarios holds a profound significance within the realm of theoretical physics. By delving into this exploration, we gain profound insights into the fundamental aspects of spacetime and Lorentz symmetry breaking. In this regard, the determination of time delay can be accomplished by rewriting the Lagrangian for null geodesics using Eq. (\ref{veff2}) and Eq. (\ref{veffr}). This mathematical approach allows us to accurately quantify the temporal delay experienced by particles, shedding light on the intricate dynamics of spacetime as follows

	\begin{equation}
	\left(\frac{{\mathrm{d}r}}{{\mathrm{d}t}}\right)^2 =f^{2}(r) \frac{1-\frac{X}{4}}{1+\frac{3 X}{4}} \left(1-\frac{ r_{\min }^2 f(r)}{r^2 f\left(r_{\min }\right)}\right)
	\end{equation}

	Moreover, we know that $\mathrm{d}r/\mathrm{d}t=0$ at $r=r_{min}$, resulting in
	\begin{equation}
	\frac{{\mathrm{d}t}}{{\mathrm{d}r}} = \frac{\sqrt{\frac{1+\frac{3X}{4}}{1-\frac{X}{4}}}}{{{f(r) }\sqrt {1 - {{(\frac{{{r_{\min }}}}{r})}^2}\frac{{{f }(r)}}{{{f }({r_{\min }})}}} }}.
	\end{equation}

In this way, the time delay is given by
\ie
t(r,r_{\text{min}}) = \int^{\infty}_{r_{\text{min}}} \frac{\sqrt{\frac{1+\frac{3X}{4}}{1-\frac{X}{4}}}}{{{f(r) }\sqrt {1 - {{(\frac{{{r_{\min }}}}{r})}^2}\frac{{{f }(r)}}{{{f }({r_{\min }})}}} }}\,dr .
\fe

		Moreover, the deflection of light as it passes through curved spacetime serves as a tool to analyze the physics of a source of gravitation \cite{bin2011strong,kumaran2023shadow,nakajima2012deflection,jusufi2018gravitational}. The calculation of the deflection angle can be carried out by applying the following equation \cite{weinberg1972gravitation}\\
	\begin{equation}\label{def}
		\hat{\alpha}= 2\int^{\infty}_{r_{\text{min}}} \left({\sqrt{\frac{\left(1-\frac{X}{4}\right)^3}{\frac{3 X}{4}+1}} \left(\frac{r^2 }{r_{\text{min}}^2}f(r_{\text{min}})-f(r)\right)}\right)^{\frac{-1}{2}} \frac{\mathrm{d}{r}}{r}-\pi.
	\end{equation}

Within the scope of our analysis, we employ the symbol $\hat{\alpha}$ to represent the deflection angle. Table. \ref{Table:timedef} provides a systematic evaluation of the time delay and deflection angle for a range of $X$ values. Our method involves initially determining the minimum radius along the trajectory for each specific $X$.

To quantify the time delay, we establish a reference point using $t_0$ as the baseline for the time delay, which corresponds to the original Schwarzschild black hole scenario with $X=0$. We then calculate the time delay ratio for each parameter within the bumblebee theory. The deflection angles are similarly derived through the application of Eq. (\ref{def}).

The results in Table. \ref{Table:timedef} reveal that while the minimum radius diminishes with increasing $X$, it is important to note that the time delay ratio, the actual time delay, and the deflection angle demonstrate distinct behaviors. Specifically, an increase in the value of $X$ is associated with longer time delays and larger deflection angles, underscoring the heightened gravitational effect corresponding to higher values of $X$.

\begin{table}[!ht]
	\centering
	\caption{Variation of $r_{min}$, time delay ratio and deflection angle for various $X$, $M=1$ and impact parameter ($L/E$) $=10$}
	\begin{tabular}{|c|c|c|c|c|c|c|c|c|c|c|c|}
		\hline
		$X$ & 0.0 & 0.1 & 0.2 & 0.3 & 0.4 & 0.5 & 0.6 & 0.7 & 0.8 & 0.9 & 1.0 \\ \hline
		$r_{min}$ & 8.7889 & 8.6684 & 8.5632 & 8.4713 & 8.3914 & 8.3222 & 8.2627 & 8.2123 & 8.1702 & 8.1359 & 8.1092 \\ \hline
		$\frac{t}{t_0}$ & 1.0000 & 1.0500 & 1.1002 & 1.1508 & 1.2019 & 1.2536 & 1.3061 & 1.3596 & 1.4142 & 1.4701 & 1.5275 \\ \hline
		$\hat{\alpha}$ & 1.8660 & 1.9421 & 2.0192 & 2.0977 & 2.1778 & 2.2599 & 2.3442 & 2.4311 & 2.5209 & 2.6139 & 2.7105 \\ \hline
	\end{tabular}
	\label{Table:timedef}
\end{table}



\section{Conclusion\label{Conc}}

This study delved into the intriguing realm of bumblebee gravity, employing the metric--affine formalism to explore its various signatures. By investigating the effects of the Lorentz violation parameter, $X$, we observed significant modifications in the renowned \textit{Hawking} temperature. Interestingly, as $X$ increased, the values of the \textit{Hawking} temperature experienced a notable attenuation.

Our analysis extended to the examination of \textit{quasinormal} modes through the utilization of the WKB method. Notably, we found that a stronger Lorentz--violating parameter led to a deceleration in the damping oscillations of gravitational waves, showcasing the intricate interplay between bumblebee gravity and Lorentz violation. In addition, we calculated the transmission and reflection coefficients in this context.

To better comprehend the implications of the \textit{quasinormal} spectrum on time--dependent scattering phenomena, we conducted a meticulous exploration of scalar perturbations in the time--domain solution. This analysis provided a comprehensive understanding of the impact and behavior of scalar perturbations under varying conditions.

Furthermore, our investigation extended to the concept of shadows, where we discovered a compelling relationship between the Lorentz violation parameter $X$ and shadow radii. Specifically, larger values of $X$ were found to correspond to increased shadow radii. More so, they were bounded from above, i.e., $X=0.86$. Lastly, our research delved into the realm of time delay and deflection angle within this scenario, shedding light on additional aspects of the intricate interplay between bumblebee gravity and its implications.





\section*{Acknowledgements}

The authors express their gratitude to the diligent referees who meticulously reviewed the manuscript and provided invaluable feedback to aid in its improvement.
Most of the calculations were performed by using the \textit{Mathematica} software. A. A. Araújo Filho is supported by Conselho Nacional de Desenvolvimento Cient\'{\i}fico e Tecnol\'{o}gico (CNPq) and Fundação de Apoio à Pesquisa do Estado da Paraíba (FAPESQ) -- [200486/2022-5] and [150891/2023-7]. The authors would like to thank R. Oliveira, R. Konoplya for providing the \textit{Mathematica} notebook to perform our numerical calculations, and P. J. Porfírio for the fruitiful discussions.

\section{Data Availability Statement}

Data Availability Statement: No Data associated in the manuscript


\bibliographystyle{ieeetr}
\bibliography{main}

\begin{thebibliography}{100}

\bibitem{1}
M.~Chaichian, A.~Tureanu, and G.~Zet, ``Corrections to schwarzschild solution
  in noncommutative gauge theory of gravity,'' {\em Physics Letters B},
  vol.~660, no.~5, pp.~573--578, 2008.

\bibitem{2}
G.~Zet, V.~Manta, and S.~Babeti, ``Desitter gauge theory of gravitation,'' {\em
  International Journal of Modern Physics C}, vol.~14, no.~01, pp.~41--48,
  2003.

\bibitem{3}
N.~Seiberg and E.~Witten, ``String theory and noncommutative geometry,'' {\em
  Journal of High Energy Physics}, vol.~1999, no.~09, p.~032, 1999.

\bibitem{4}
A.~H. Chamseddine, ``Deforming einstein's gravity,'' {\em Physics Letters B},
  vol.~504, no.~1-2, pp.~33--37, 2001.

\bibitem{5}
B.~Jurco, S.~Schraml, P.~Schupp, and J.~Wess, ``Enveloping algebra-valued gauge
  transformations for non-abelian gauge groups on non-commutative spaces,''
  {\em The European Physical Journal C-Particles and Fields}, vol.~17, no.~3,
  pp.~521--526, 2000.

\bibitem{6}
F.~W. Hehl, P.~Von~der Heyde, G.~D. Kerlick, and J.~M. Nester, ``General
  relativity with spin and torsion: Foundations and prospects,'' {\em Reviews
  of Modern Physics}, vol.~48, no.~3, p.~393, 1976.

\bibitem{7}
R.-G. Cai and K.-S. Soh, ``Topological black holes in the dimensionally
  continued gravity,'' {\em Physical Review D}, vol.~59, no.~4, p.~044013,
  1999.

\bibitem{8}
L.~Xiang and X.~Wen, ``A heuristic analysis of black hole thermodynamics with
  generalized uncertainty principle,'' {\em Journal of High Energy Physics},
  vol.~2009, no.~10, p.~046, 2009.

\bibitem{9}
D.~Kastor, S.~Ray, and J.~Traschen, ``Enthalpy and the mechanics of ads black
  holes,'' {\em Classical and Quantum Gravity}, vol.~26, no.~19, p.~195011,
  2009.

\bibitem{10}
W.~Jesus and A.~Santos, ``Ricci dark energy in bumblebee gravity model,'' {\em
  Modern Physics Letters A}, vol.~34, no.~22, p.~1950171, 2019.

\bibitem{11}
R.~Maluf and J.~C. Neves, ``Bumblebee field as a source of cosmological
  anisotropies,'' {\em Journal of Cosmology and Astroparticle Physics},
  vol.~2021, no.~10, p.~038, 2021.

\bibitem{12}
R.~Maluf and J.~C. Neves, ``Black holes with a cosmological constant in
  bumblebee gravity,'' {\em Physical Review D}, vol.~103, no.~4, p.~044002,
  2021.

\bibitem{13}
S.~K. Jha, H.~Barman, and A.~Rahaman, ``Bumblebee gravity and particle motion
  in snyder noncommutative spacetime structures,'' {\em Journal of Cosmology
  and Astroparticle Physics}, vol.~2021, no.~04, p.~036, 2021.

\bibitem{14}
R.~Jackiw and S.-Y. Pi, ``Chern-simons modification of general relativity,''
  {\em Physical Review D}, vol.~68, no.~10, p.~104012, 2003.

\bibitem{15}
L.~Mirzagholi, E.~Komatsu, K.~D. Lozanov, and Y.~Watanabe, ``Effects of
  gravitational chern-simons during axion-su (2) inflation,'' {\em Journal of
  Cosmology and Astroparticle Physics}, vol.~2020, no.~06, p.~024, 2020.

\bibitem{16}
N.~Bartolo and G.~Orlando, ``Parity breaking signatures from a chern-simons
  coupling during inflation: the case of non-gaussian gravitational waves,''
  {\em Journal of Cosmology and Astroparticle Physics}, vol.~2017, no.~07,
  p.~034, 2017.

\bibitem{17}
A.~Conroy and T.~Koivisto, ``Parity-violating gravity and gw170817 in
  non-riemannian cosmology,'' {\em Journal of Cosmology and Astroparticle
  Physics}, vol.~2019, no.~12, p.~016, 2019.

\bibitem{18}
M.~Li, H.~Rao, and D.~Zhao, ``A simple parity violating gravity model without
  ghost instability,'' {\em Journal of Cosmology and Astroparticle Physics},
  vol.~2020, no.~11, p.~023, 2020.

\bibitem{19}
P.~Porfirio, J.~Fonseca-Neto, J.~Nascimento, A.~Y. Petrov, J.~Ricardo, and
  A.~Santos, ``Chern-simons modified gravity and closed timelike curves,'' {\em
  Physical Review D}, vol.~94, no.~4, p.~044044, 2016.

\bibitem{20}
P.~Porfirio, J.~Fonseca-Neto, J.~Nascimento, and A.~Y. Petrov, ``Causality
  aspects of the dynamical chern-simons modified gravity,'' {\em Physical
  Review D}, vol.~94, no.~10, p.~104057, 2016.

\bibitem{21}
B.~Altschul, J.~Nascimento, A.~Y. Petrov, and P.~Porf{\'\i}rio, ``First-order
  perturbations of g{\"o}del-type metrics in non-dynamical chern--simons
  modified gravity,'' {\em Classical and Quantum Gravity}, vol.~39, no.~2,
  p.~025002, 2021.

\bibitem{22}
J.~Nascimento, A.~Y. Petrov, and P.~Porf{\'\i}rio, ``Induced gravitational
  topological term and the einstein-cartan modified theory,'' {\em Physical
  Review D}, vol.~105, no.~4, p.~044053, 2022.

\bibitem{23}
D.~Bao, S.-S. Chern, and Z.~Shen, {\em An introduction to Riemann-Finsler
  geometry}, vol.~200.
\newblock Springer Science \& Business Media, 2000.

\bibitem{24}
J.~Foster and R.~Lehnert, ``Classical-physics applications for finsler b
  space,'' {\em Physics Letters B}, vol.~746, pp.~164--170, 2015.

\bibitem{25}
B.~R. Edwards and V.~A. Kosteleck{\`y}, ``Riemann--finsler geometry and
  lorentz-violating scalar fields,'' {\em Physics Letters B}, vol.~786,
  pp.~319--326, 2018.

\bibitem{26}
M.~Schreck, ``Classical kinematics and finsler structures for nonminimal
  lorentz-violating fermions,'' {\em The European Physical Journal C}, vol.~75,
  pp.~1--16, 2015.

\bibitem{27}
D.~Colladay and P.~McDonald, ``Singular lorentz-violating lagrangians and
  associated finsler structures,'' {\em Physical Review D}, vol.~92, no.~8,
  p.~085031, 2015.

\bibitem{28}
M.~Schreck, ``Classical lagrangians and finsler structures for the nonminimal
  fermion sector of the standard model extension,'' {\em Physical Review D},
  vol.~93, no.~10, p.~105017, 2016.

\bibitem{31}
A.~Delhom, J.~Nascimento, G.~J. Olmo, A.~Y. Petrov, and P.~J. Porf{\'\i}rio,
  ``Metric-affine bumblebee gravity: classical aspects,'' {\em The European
  Physical Journal C}, vol.~81, pp.~1--10, 2021.

\bibitem{32}
A.~Delhom, J.~Nascimento, G.~J. Olmo, A.~Y. Petrov, and P.~J. Porf{\'\i}rio,
  ``Radiative corrections in metric-affine bumblebee model,'' {\em Physics
  Letters B}, vol.~826, p.~136932, 2022.

\bibitem{33}
A.~Delhom, T.~Mariz, J.~Nascimento, G.~J. Olmo, A.~Y. Petrov, and P.~J.
  Porf{\'\i}rio, ``Spontaneous lorentz symmetry breaking and one-loop effective
  action in the metric-affine bumblebee gravity,'' {\em Journal of Cosmology
  and Astroparticle Physics}, vol.~2022, no.~07, p.~018, 2022.

\bibitem{34}
S.~Boudet, F.~Bombacigno, G.~J. Olmo, and P.~J. Porfirio, ``Quasinormal modes
  of schwarzschild black holes in projective invariant chern-simons modified
  gravity,'' {\em Journal of Cosmology and Astroparticle Physics}, vol.~2022,
  no.~05, p.~032, 2022.

\bibitem{nascimento2022vacuum}
A.~A. Ara{\'u}jo~Filho, J.~Nascimento, A.~Y. Petrov, and P.~J. Porf{\'\i}rio,
  ``Vacuum solution within a metric-affine bumblebee gravity,'' {\em Physical
  Review D}, vol.~108, no.~8, p.~085010, 2023.

\bibitem{unno1979nonradial}
W.~Unno, Y.~Osaki, H.~Ando, and H.~Shibahashi, ``Nonradial oscillations of
  stars,'' {\em Tokyo: University of Tokyo Press}, 1979.

\bibitem{kjeldsen1994amplitudes}
H.~Kjeldsen and T.~R. Bedding, ``Amplitudes of stellar oscillations: the
  implications for asteroseismology,'' {\em arXiv preprint astro-ph/9403015},
  1994.

\bibitem{dziembowski1992effects}
W.~Dziembowski and P.~R. Goode, ``Effects of differential rotation on stellar
  oscillations-a second-order theory,'' {\em The Astrophysical Journal},
  vol.~394, pp.~670--687, 1992.

\bibitem{pretorius2005evolution}
F.~Pretorius, ``Evolution of binary black-hole spacetimes,'' {\em Physical
  review letters}, vol.~95, no.~12, p.~121101, 2005.

\bibitem{hurley2002evolution}
J.~R. Hurley, C.~A. Tout, and O.~R. Pols, ``Evolution of binary stars and the
  effect of tides on binary populations,'' {\em Monthly Notices of the Royal
  Astronomical Society}, vol.~329, no.~4, pp.~897--928, 2002.

\bibitem{yakut2005evolution}
K.~Yakut and P.~P. Eggleton, ``Evolution of close binary systems,'' {\em The
  Astrophysical Journal}, vol.~629, no.~2, p.~1055, 2005.

\bibitem{heuvel2011compact}
E.~v.~d. Heuvel, ``Compact stars and the evolution of binary systems,'' in {\em
  Fluid Flows To Black Holes: A Tribute to S Chandrasekhar on His Birth
  Centenary}, pp.~55--73, World Scientific, 2011.

\bibitem{riles2017recent}
K.~Riles, ``Recent searches for continuous gravitational waves,'' {\em Modern
  Physics Letters A}, vol.~32, no.~39, p.~1730035, 2017.

\bibitem{konoplya2011quasinormal}
R.~Konoplya and A.~Zhidenko, ``Quasinormal modes of black holes: From
  astrophysics to string theory,'' {\em Reviews of Modern Physics}, vol.~83,
  no.~3, p.~793, 2011.

\bibitem{heidari2023gravitational}
N.~Heidari, H.~Hassanabadi, J.~Kur{\'\i}uz, S.~Zare, P.~Porf{\'\i}rio, {\em
  et~al.}, ``Gravitational signatures of a non--commutative stable black
  hole,'' {\em arXiv preprint arXiv:2305.06838}, 2023.

\bibitem{kokkotas1999quasi}
K.~D. Kokkotas and B.~G. Schmidt, ``Quasi-normal modes of stars and black
  holes,'' {\em Living Reviews in Relativity}, vol.~2, no.~1, pp.~1--72, 1999.

\bibitem{rincon2020greybody}
{\'A}.~Rinc{\'o}n and V.~Santos, ``Greybody factor and quasinormal modes of
  regular black holes,'' {\em The European Physical Journal C}, vol.~80,
  no.~10, pp.~1--7, 2020.

\bibitem{santos2016quasinormal}
V.~Santos, R.~Maluf, and C.~Almeida, ``Quasinormal frequencies of self-dual
  black holes,'' {\em Physical Review D}, vol.~93, no.~8, p.~084047, 2016.

\bibitem{oliveira2019quasinormal}
R.~Oliveira, D.~Dantas, V.~Santos, and C.~Almeida, ``Quasinormal modes of
  bumblebee wormhole,'' {\em Classical and Quantum Gravity}, vol.~36, no.~10,
  p.~105013, 2019.

\bibitem{berti2009quasinormal}
E.~Berti, V.~Cardoso, and A.~O. Starinets, ``Quasinormal modes of black holes
  and black branes,'' {\em Classical and Quantum Gravity}, vol.~26, no.~16,
  p.~163001, 2009.

\bibitem{horowitz2000quasinormal}
G.~T. Horowitz and V.~E. Hubeny, ``Quasinormal modes of ads black holes and the
  approach to thermal equilibrium,'' {\em Physical Review D}, vol.~62, no.~2,
  p.~024027, 2000.

\bibitem{nollert1999quasinormal}
H.-P. Nollert, ``Quasinormal modes: the characteristicsound'of black holes and
  neutron stars,'' {\em Classical and Quantum Gravity}, vol.~16, no.~12,
  p.~R159, 1999.

\bibitem{ferrari1984new}
V.~Ferrari and B.~Mashhoon, ``New approach to the quasinormal modes of a black
  hole,'' {\em Physical Review D}, vol.~30, no.~2, p.~295, 1984.

\bibitem{london2014modeling}
L.~London, D.~Shoemaker, and J.~Healy, ``Modeling ringdown: Beyond the
  fundamental quasinormal modes,'' {\em Physical Review D}, vol.~90, no.~12,
  p.~124032, 2014.

\bibitem{maggiore2008physical}
M.~Maggiore, ``Physical interpretation of the spectrum of black hole
  quasinormal modes,'' {\em Physical Review Letters}, vol.~100, no.~14,
  p.~141301, 2008.

\bibitem{flachi2013quasinormal}
A.~Flachi and J.~P. Lemos, ``Quasinormal modes of regular black holes,'' {\em
  Physical Review D}, vol.~87, no.~2, p.~024034, 2013.

\bibitem{ovgun2018quasinormal}
A.~{\"O}vg{\"u}n, I.~Sakall{\i}, and J.~Saavedra, ``Quasinormal modes of a
  schwarzschild black hole immersed in an electromagnetic universe,'' {\em
  Chinese Physics C}, vol.~42, no.~10, p.~105102, 2018.

\bibitem{blazquez2018scalar}
J.~L. Bl{\'a}zquez-Salcedo, X.~Y. Chew, and J.~Kunz, ``Scalar and axial
  quasinormal modes of massive static phantom wormholes,'' {\em Physical Review
  D}, vol.~98, no.~4, p.~044035, 2018.

\bibitem{roy2020revisiting}
P.~D. Roy, S.~Aneesh, and S.~Kar, ``Revisiting a family of wormholes: geometry,
  matter, scalar quasinormal modes and echoes,'' {\em The European Physical
  Journal C}, vol.~80, no.~9, pp.~1--17, 2020.

\bibitem{kim2018quasi}
J.~Y. Kim, C.~O. Lee, and M.-I. Park, ``Quasi-normal modes of a natural ads
  wormhole in einstein--born--infeld gravity,'' {\em The European Physical
  Journal C}, vol.~78, no.~12, pp.~1--15, 2018.

\bibitem{lee2020quasi}
C.~O. Lee, J.~Y. Kim, and M.-I. Park, ``Quasi-normal modes and stability of
  einstein--born--infeld black holes in de sitter space,'' {\em The European
  Physical Journal C}, vol.~80, no.~8, pp.~1--21, 2020.

\bibitem{jawad2020quasinormal}
A.~Jawad, S.~Chaudhary, M.~Yasir, A.~{\"O}vg{\"u}n, and {\.I}.~Sakall{\i},
  ``Quasinormal modes of extended gravity black holes,'' 2020.

\bibitem{maluf2013matter}
R.~Maluf, V.~Santos, W.~Cruz, and C.~Almeida, ``Matter-gravity scattering in
  the presence of spontaneous lorentz violation,'' {\em Physical Review D},
  vol.~88, no.~2, p.~025005, 2013.

\bibitem{maluf2014einstein}
R.~Maluf, C.~Almeida, R.~Casana, and M.~Ferreira~Jr, ``Einstein-hilbert
  graviton modes modified by the lorentz-violating bumblebee field,'' {\em
  Physical Review D}, vol.~90, no.~2, p.~025007, 2014.

\bibitem{JCAP1}
M.~Okyay and A.~{\"O}vg{\"u}n, ``Nonlinear electrodynamics effects on the black
  hole shadow, deflection angle, quasinormal modes and greybody factors,'' {\em
  Journal of Cosmology and Astroparticle Physics}, vol.~2022, no.~01, p.~009,
  2022.

\bibitem{JCAP2}
Y.~Zhao, X.~Ren, A.~Ilyas, E.~N. Saridakis, and Y.-F. Cai, ``Quasinormal modes
  of black holes in f (t) gravity,'' {\em Journal of Cosmology and
  Astroparticle Physics}, vol.~2022, no.~10, p.~087, 2022.

\bibitem{JCAP3}
S.~Boudet, F.~Bombacigno, G.~J. Olmo, and P.~J. Porfirio, ``Quasinormal modes
  of schwarzschild black holes in projective invariant chern-simons modified
  gravity,'' {\em Journal of Cosmology and Astroparticle Physics}, vol.~2022,
  no.~05, p.~032, 2022.

\bibitem{jcap4}
M.~Cadoni, M.~Oi, and A.~P. Sanna, ``Quasi-normal modes and microscopic
  description of 2d black holes,'' {\em Journal of High Energy Physics},
  vol.~2022, no.~1, pp.~1--23, 2022.

\bibitem{jcap5}
L.~Hui, D.~Kabat, and S.~S. Wong, ``Quasinormal modes, echoes and the causal
  structure of the green's function,'' {\em Journal of Cosmology and
  Astroparticle Physics}, vol.~2019, no.~12, p.~020, 2019.

\bibitem{abbott2016ligo}
B.~Abbott, S.~Jawahar, N.~Lockerbie, and K.~Tokmakov, ``Ligo scientific
  collaboration and virgo collaboration (2016) directly comparing gw150914 with
  numerical solutions of einstein's equations for binary black hole
  coalescence. physical review d, 94 (6). issn 1550-2368, http://dx. doi.
  org/10.1103/physrevd. 94.064035,'' {\em PHYSICAL REVIEW D Phys Rev D},
  vol.~94, p.~064035, 2016.

\bibitem{abbott2017gravitational}
B.~P. Abbott, R.~Abbott, T.~Abbott, F.~Acernese, K.~Ackley, C.~Adams, T.~Adams,
  P.~Addesso, R.~Adhikari, V.~Adya, {\em et~al.}, ``Gravitational waves and
  gamma-rays from a binary neutron star merger: Gw170817 and grb 170817a,''
  {\em The Astrophysical Journal Letters}, vol.~848, no.~2, p.~L13, 2017.

\bibitem{abbott2017gw170817}
B.~P. Abbott, R.~Abbott, T.~Abbott, F.~Acernese, K.~Ackley, C.~Adams, T.~Adams,
  P.~Addesso, R.~Adhikari, V.~Adya, {\em et~al.}, ``Gw170817: observation of
  gravitational waves from a binary neutron star inspiral,'' {\em Physical
  Review Letters}, vol.~119, no.~16, p.~161101, 2017.

\bibitem{abbott2017multi}
B.~P. Abbott, S.~Bloemen, P.~Canizares, H.~Falcke, R.~Fender, S.~Ghosh,
  P.~Groot, T.~Hinderer, J.~H{\"o}randel, P.~Jonker, {\em et~al.},
  ``Multi-messenger observations of a binary neutron star merger,'' 2017.

\bibitem{fafone2015advanced}
V.~Fafone, ``Advanced virgo: an update,'' in {\em THE THIRTEENTH MARCEL
  GROSSMANN MEETING: On Recent Developments in Theoretical and Experimental
  General Relativity, Astrophysics and Relativistic Field Theories},
  pp.~2025--2028, World Scientific, 2015.

\bibitem{abramovici1992ligo}
A.~Abramovici, W.~E. Althouse, R.~W. Drever, Y.~G{\"u}rsel, S.~Kawamura, F.~J.
  Raab, D.~Shoemaker, L.~Sievers, R.~E. Spero, K.~S. Thorne, {\em et~al.},
  ``Ligo: The laser interferometer gravitational-wave observatory,'' {\em
  science}, vol.~256, no.~5055, pp.~325--333, 1992.

\bibitem{coccia1995gravitational}
E.~Coccia, G.~Pizzella, and F.~Ronga, {\em Gravitational Wave
  Experiments-Proceedings Of The First Edoardo Amaldi Conference}, vol.~1.
\newblock World Scientific, 1995.

\bibitem{luck1997geo600}
H.~L{\"u}ck, G.~Team, {\em et~al.}, ``The geo600 project,'' {\em Classical and
  quantum gravity}, vol.~14, no.~6, p.~1471, 1997.

\bibitem{evans2014gravitational}
M.~Evans, ``Gravitational wave detection with advanced ground based
  detectors,'' {\em General Relativity and Gravitation}, vol.~46, no.~10,
  p.~1778, 2014.

\bibitem{thorne2000probing}
K.~S. Thorne, ``Probing black holes and relativistic stars with gravitational
  waves,'' in {\em Black Holes and the Structure of the Universe}, pp.~81--118,
  World Scientific, 2000.

\bibitem{regge1957stability}
T.~Regge and J.~A. Wheeler, ``Stability of a schwarzschild singularity,'' {\em
  Physical Review}, vol.~108, no.~4, p.~1063, 1957.

\bibitem{zerilli1970effective}
F.~J. Zerilli, ``Effective potential for even-parity regge-wheeler
  gravitational perturbation equations,'' {\em Physical Review Letters},
  vol.~24, no.~13, p.~737, 1970.

\bibitem{zerilli1974perturbation}
F.~J. Zerilli, ``Perturbation analysis for gravitational and electromagnetic
  radiation in a reissner-nordstr{\"o}m geometry,'' {\em Physical Review D},
  vol.~9, no.~4, p.~860, 1974.

\bibitem{herdeiro2015asymptotically}
C.~A. Herdeiro and E.~Radu, ``Asymptotically flat black holes with scalar hair:
  a review,'' {\em International Journal of Modern Physics D}, vol.~24, no.~09,
  p.~1542014, 2015.

\bibitem{ayon2016analytic}
E.~Ay{\'o}n-Beato, F.~Canfora, and J.~Zanelli, ``Analytic self-gravitating
  skyrmions, cosmological bounces and ads wormholes,'' {\em Physics Letters B},
  vol.~752, pp.~201--205, 2016.

\bibitem{colpi1986boson}
M.~Colpi, S.~L. Shapiro, and I.~Wasserman, ``Boson stars: gravitational
  equilibria of self-interacting scalar fields,'' {\em Physical review
  letters}, vol.~57, no.~20, p.~2485, 1986.

\bibitem{palenzuela2017gravitational}
C.~Palenzuela, P.~Pani, M.~Bezares, V.~Cardoso, L.~Lehner, and S.~Liebling,
  ``Gravitational wave signatures of highly compact boson star binaries,'' {\em
  Physical Review D}, vol.~96, no.~10, p.~104058, 2017.

\bibitem{cunha2017lensing}
P.~V. Cunha, J.~A. Font, C.~Herdeiro, E.~Radu, N.~Sanchis-Gual, and M.~Zilhao,
  ``Lensing and dynamics of ultracompact bosonic stars,'' {\em Physical Review
  D}, vol.~96, no.~10, p.~104040, 2017.

\bibitem{visser2004stable}
M.~Visser and D.~L. Wiltshire, ``Stable gravastars—an alternative to black
  holes?,'' {\em Classical and Quantum Gravity}, vol.~21, no.~4, p.~1135, 2004.

\bibitem{pani2009gravitational}
P.~Pani, E.~Berti, V.~Cardoso, Y.~Chen, and R.~Norte, ``Gravitational wave
  signatures of the absence of an event horizon: Nonradial oscillations of a
  thin-shell gravastar,'' {\em Physical Review D}, vol.~80, no.~12, p.~124047,
  2009.

\bibitem{chirenti2016did}
C.~Chirenti and L.~Rezzolla, ``Did gw150914 produce a rotating gravastar?,''
  {\em Physical Review D}, vol.~94, no.~8, p.~084016, 2016.

\bibitem{cardoso2004black}
V.~Cardoso, O.~J. Dias, J.~P. Lemos, and S.~Yoshida, ``Black-hole bomb and
  superradiant instabilities,'' {\em Physical Review D}, vol.~70, no.~4,
  p.~044039, 2004.

\bibitem{sanchis2016explosion}
N.~Sanchis-Gual, J.~C. Degollado, P.~J. Montero, J.~A. Font, and C.~Herdeiro,
  ``Explosion and final state of an unstable reissner-nordstr{\"o}m black
  hole,'' {\em Physical review letters}, vol.~116, no.~14, p.~141101, 2016.

\bibitem{hod2016charged}
S.~Hod, ``The charged black-hole bomb: A lower bound on the charge-to-mass
  ratio of the explosive scalar field,'' {\em Physics Letters B}, vol.~755,
  pp.~177--182, 2016.

\bibitem{brito2015black}
R.~Brito, V.~Cardoso, and P.~Pani, ``Black holes as particle detectors:
  evolution of superradiant instabilities,'' {\em Classical and Quantum
  Gravity}, vol.~32, no.~13, p.~134001, 2015.

\bibitem{35}
Y.~Bonder, ``Lorentz violation in the gravity sector: The t puzzle,'' {\em
  Physical Review D}, vol.~91, no.~12, p.~125002, 2015.

\bibitem{maluf2019antisymmetric}
R.~Maluf, A.~Ara{\'u}jo~Filho, W.~Cruz, and C.~Almeida, ``Antisymmetric tensor
  propagator with spontaneous lorentz violation,'' {\em Europhysics Letters},
  vol.~124, no.~6, p.~61001, 2019.

\bibitem{36}
V.~I. Alfonso, C.~Bejarano, J.~B. Jimenez, G.~J. Olmo, and E.~Orazi, ``The
  trivial role of torsion in projective invariant theories of gravity with
  non-minimally coupled matter fields,'' {\em Classical and Quantum Gravity},
  vol.~34, no.~23, p.~235003, 2017.

\bibitem{37}
J.~B. Jim{\'e}nez, L.~Heisenberg, G.~J. Olmo, and D.~Rubiera-Garcia,
  ``Born--infeld inspired modifications of gravity,'' {\em Physics Reports},
  vol.~727, pp.~1--129, 2018.

\bibitem{bekenstein1973black}
J.~D. Bekenstein, ``Black holes and entropy,'' {\em Physical Review D}, vol.~7,
  no.~8, p.~2333, 1973.

\bibitem{bekenstein1974generalized}
J.~D. Bekenstein, ``Generalized second law of thermodynamics in black-hole
  physics,'' {\em Physical Review D}, vol.~9, no.~12, p.~3292, 1974.

\bibitem{bekenstein2020black}
J.~D. Bekenstein, ``Black holes and the second law,'' in {\em JACOB BEKENSTEIN:
  The Conservative Revolutionary}, pp.~303--306, World Scientific, 2020.

\bibitem{bardeen1973four}
J.~M. Bardeen, B.~Carter, and S.~W. Hawking, ``The four laws of black hole
  mechanics,'' {\em Communications in mathematical physics}, vol.~31,
  pp.~161--170, 1973.

\bibitem{jacobson1995thermodynamics}
T.~Jacobson, ``Thermodynamics of spacetime: the einstein equation of state,''
  {\em Physical Review Letters}, vol.~75, no.~7, p.~1260, 1995.

\bibitem{padmanabhan2010thermodynamical}
T.~Padmanabhan, ``Thermodynamical aspects of gravity: new insights,'' {\em
  Reports on Progress in Physics}, vol.~73, no.~4, p.~046901, 2010.

\bibitem{araujo2021bouncing}
A.~A. Ara{\'u}jo~Filho and A.~Y. Petrov, ``Bouncing universe in a heat bath,''
  {\em International Journal of Modern Physics A}, vol.~36, p.~2150242, 2021.

\bibitem{araujo2022thermal}
A.~A. Ara{\'u}jo~Filho, {\em Thermal aspects of field theories}.
\newblock Amazon. com, 2022.

\bibitem{araujo2023thermodynamics}
A.~A. Ara{\'u}jo~Filho, S.~Zare, P.~Porf{\'\i}rio, J.~K{\v{r}}{\'\i}{\v{z}},
  and H.~Hassanabadi, ``Thermodynamics and evaporation of a modified
  schwarzschild black hole in a non--commutative gauge theory,'' {\em Physics
  Letters B}, vol.~838, p.~137744, 2023.

\bibitem{aa2}
A.~A. Ara{\'u}jo~Filho and J.~Reis, ``Thermal aspects of interacting quantum
  gases in lorentz-violating scenarios,'' {\em The European Physical Journal
  Plus}, vol.~136, pp.~1--30, 2021.

\bibitem{aa4}
A.~A. Ara{\'u}jo~Filho, ``Lorentz-violating scenarios in a thermal reservoir,''
  {\em The European Physical Journal Plus}, vol.~136, no.~4, pp.~1--14, 2021.

\bibitem{aa6}
A.~A. Ara{\'u}jo~Filho and R.~V. Maluf, ``Thermodynamic properties in
  higher-derivative electrodynamics,'' {\em Brazilian Journal of Physics},
  vol.~51, pp.~820--830, 2021.

\bibitem{aa7}
A.~A. Ara{\'u}jo~Filho and A.~Y. Petrov, ``Higher-derivative lorentz-breaking
  dispersion relations: a thermal description,'' {\em The European Physical
  Journal C}, vol.~81, no.~9, p.~843, 2021.

\bibitem{aa10}
A.~A. Ara{\'u}jo~Filho, ``Particles in loop quantum gravity formalism: a
  thermodynamical description,'' {\em Annalen der Physik}, p.~2200383, 2022.

\bibitem{aa13}
P.~Sedaghatnia, H.~Hassanabadi, J.~Porf{\'\i}rio, W.~Chung, {\em et~al.},
  ``Thermodynamical properties of a deformed schwarzschild black hole via dunkl
  generalization,'' {\em arXiv preprint arXiv:2302.11460}, 2023.

\bibitem{aa14}
A.~A. Ara{\'u}jo~Filho, J.~Furtado, and J.~Silva, ``Thermodynamical properties
  of an ideal gas in a traversable wormhole,'' {\em arXiv preprint
  arXiv:2302.05492}, 2023.

\bibitem{aa15}
J.~Furtado, H.~Hassanabadi, J.~Reis, {\em et~al.}, ``Thermal analysis of
  photon-like particles in rainbow gravity,'' {\em arXiv preprint
  arXiv:2305.08587}, 2023.

\bibitem{filho2022thermodynamics}
A.~A. Ara{\'u}jo~Filho, ``Thermodynamics of massless particles in curved
  spacetime,'' {\em arXiv preprint arXiv:2201.00066}, 2022.

\bibitem{aa11}
A.~A. Ara{\'u}jo~Filho, J.~Reis, and S.~Ghosh, ``Fermions on a torus knot,''
  {\em The European Physical Journal Plus}, vol.~137, no.~5, p.~614, 2022.

\bibitem{aa12}
A.~A. Ara{\'u}jo~Filho and J.~Reis, ``How does geometry affect quantum
  gases?,'' {\em International Journal of Modern Physics A}, vol.~37,
  no.~11n12, p.~2250071, 2022.

\bibitem{aa1}
R.~R. Oliveira, A.~A. Ara{\'u}jo~Filho, F.~C. Lima, R.~V. Maluf, and C.~A.
  Almeida, ``Thermodynamic properties of an aharonov-bohm quantum ring,'' {\em
  The European Physical Journal Plus}, vol.~134, no.~10, p.~495, 2019.

\bibitem{qq}
A.~A. Ara{\'u}jo~Filho, H.~Hassanabadi, J.~A. A. d. S.~d. Reis, and L.~L.
  Santos, ``Thermodynamics of a quantum ring modified by lorentz violation,''
  {\em Physica Scripta}, 2022.

\bibitem{iyer1987black}
S.~Iyer and C.~M. Will, ``Black-hole normal modes: A wkb approach. i.
  foundations and application of a higher-order wkb analysis of
  potential-barrier scattering,'' {\em Physical Review D}, vol.~35, no.~12,
  p.~3621, 1987.

\bibitem{iyer1987black1}
S.~Iyer, ``Black-hole normal modes: A wkb approach. ii. schwarzschild black
  holes,'' {\em Physical Review D}, vol.~35, no.~12, p.~3632, 1987.

\bibitem{konoplya2003quasinormal}
R.~Konoplya, ``Quasinormal behavior of the d-dimensional schwarzschild black
  hole and the higher order wkb approach,'' {\em Physical Review D}, vol.~68,
  no.~2, p.~024018, 2003.

\bibitem{schutz1985black}
B.~F. Schutz and C.~M. Will, ``Black hole normal modes: a semianalytic
  approach,'' {\em The Astrophysical Journal}, vol.~291, pp.~L33--L36, 1985.

\bibitem{konoplya2004quasinormal}
R.~Konoplya, ``Quasinormal modes of the schwarzschild black hole and higher
  order wkb approach,'' {\em J. Phys. Stud}, vol.~8, p.~93, 2004.

\bibitem{matyjasek2017quasinormal}
J.~Matyjasek and M.~Opala, ``Quasinormal modes of black holes: The improved
  semianalytic approach,'' {\em Physical Review D}, vol.~96, no.~2, p.~024011,
  2017.

\bibitem{chen2023quasinormal}
H.~Chen, T.~Sathiyaraj, H.~Hassanabadi, Y.~Yang, Z.-W. Long, and F.-Q. Tu,
  ``Quasinormal modes of the egup-corrected schwarzschild black hole,'' {\em
  Indian Journal of Physics}, pp.~1--9, 2023.

\bibitem{konoplya2020quantum}
R.~Konoplya, ``Quantum corrected black holes: Quasinormal modes, scattering,
  shadows,'' {\em Physics Letters B}, vol.~804, p.~135363, 2020.

\bibitem{konoplya2019higher}
R.~Konoplya, A.~Zhidenko, and A.~Zinhailo, ``Higher order wkb formula for
  quasinormal modes and grey-body factors: recipes for quick and accurate
  calculations,'' {\em Classical and Quantum Gravity}, vol.~36, no.~15,
  p.~155002, 2019.

\bibitem{campos2022quasinormal}
J.~Campos, M.~Anacleto, F.~Brito, and E.~Passos, ``Quasinormal modes and shadow
  of noncommutative black hole,'' {\em Scientific Reports}, vol.~12, no.~1,
  p.~8516, 2022.

\bibitem{gundlach1994late}
C.~Gundlach, R.~H. Price, and J.~Pullin, ``Late-time behavior of stellar
  collapse and explosions. i. linearized perturbations,'' {\em Physical Review
  D}, vol.~49, no.~2, p.~883, 1994.

\bibitem{carter1968global}
B.~Carter, ``Global structure of the kerr family of gravitational fields,''
  {\em Physical Review}, vol.~174, no.~5, p.~1559, 1968.

\bibitem{singh2018shadow}
B.~P. Singh and S.~G. Ghosh, ``Shadow of schwarzschild--tangherlini black
  holes,'' {\em Annals of Physics}, vol.~395, pp.~127--137, 2018.

\bibitem{vagnozzi2022horizon}
S.~Vagnozzi, R.~Roy, Y.-D. Tsai, L.~Visinelli, M.~Afrin, A.~Allahyari,
  P.~Bambhaniya, D.~Dey, S.~G. Ghosh, P.~S. Joshi, {\em et~al.},
  ``Horizon-scale tests of gravity theories and fundamental physics from the
  event horizon telescope image of sagittarius a,'' {\em Classical and Quantum
  Gravity}, 2022.

\bibitem{bin2011strong}
A.~Y. Bin-Nun, ``Strong gravitational lensing by sgr a,'' {\em Classical and
  Quantum Gravity}, vol.~28, no.~11, p.~114003, 2011.

\bibitem{kumaran2023shadow}
Y.~Kumaran and A.~{\"O}vg{\"u}n, ``Shadow and deflection angle of asymptotic,
  magnetically-charged, non-singular black hole,'' {\em arXiv preprint
  arXiv:2306.04705}, 2023.

\bibitem{nakajima2012deflection}
K.~Nakajima and H.~Asada, ``Deflection angle of light in an ellis wormhole
  geometry,'' {\em Physical Review D}, vol.~85, no.~10, p.~107501, 2012.

\bibitem{jusufi2018gravitational}
K.~Jusufi and A.~{\"O}vg{\"u}n, ``Gravitational lensing by rotating
  wormholes,'' {\em Physical Review D}, vol.~97, no.~2, p.~024042, 2018.

\bibitem{weinberg1972gravitation}
S.~Weinberg, ``Gravitation and cosmology: principles and applications of the
  general theory of relativity,'' 1972.

\end{thebibliography}

\end{document}